**Manuscript Title:** Intrinsic Brain Networks Underlying the Experience and Expression of Subclinical Anxiety

**Running Title:** Dissociable Resting-State Networks of Subclinical Anxiety

**Authors:** Shruti Kinger[a], Naviya Lall[a], and Mrinmoy Chakrabarty[a]

**Affiliations:** [a]Cognitive Science Lab, Dept. of Social Sciences and Humanities, Indraprastha Institute of Information Technology Delhi, 110020, New Delhi, India

**Correspondence should be addressed to:** mrinmoy@iiitd.ac.in, Tel: +91-11-26907-363

**Number of Figures -** 4

**Number of Tables -** 3

## ABSTRACT

Anxiety encompasses behavioural, physiological, and subjective components that do not always align, yet it remains unclear whether these dimensions are supported by distinct intrinsic brain networks. Drawing on the two-system framework, we examined whether resting-state functional connectivity (rsFC) differentiates these components in subclinical anxiety. Forty-seven young adults with varying levels of subclinical anxiety completed a threat anticipation task to assess subclinical anxiety–modulated behavioural responses (reaction time) and physiological arousal (skin conductance), alongside the NIH Fear-Affect self-reported measure of anxiety severity. These measures were then related to rsFC using independent component analyses. Higher levels of subclinical anxiety were associated with faster responses during temporally uncertain threat, consistent with escape-like behaviour, whereas no relationship emerged between subclinical anxiety and physiological arousal. At the neural level, three distinct patterns of connectivity emerged that remained stable after sequential false-discovery rate correction. Subclinical anxiety–modulated behavioural responses were associated with stronger connectivity between the default mode network and regions in visual cortex and paracingulate gyrus. Physiological modulation by subclinical anxiety was linked to connectivity between the salience network and visual cortex. In contrast, subjective subclinical anxiety associated with connectivity between the cerebellar network and regions in frontal, parietal and visual cortices. Altogether, the findings suggest that behavioural, physiological, and subjective aspects of subclinical anxiety relate to functionally and spatially dissociable intrinsic brain networks, extending prior task-based findings to resting-state connectivity and informing future work on early neural markers of anxiety.

## INTRODUCTION

Anxiety is an affective experience involving apprehension, worry, tension, and anticipation of potential future threat (American Psychological Association, 2018). It is a complex construct involving interacting mental and bodily states that together shape behavioural responses, with the internal state acquiring a negative valence when anxiety becomes excessive or dysregulated (Lang, 1968; LeDoux & Pine, 2016). Although anxiety has clear adaptive value and is rooted in biological and psychological predispositions, it can become maladaptive when it is excessive in intensity, frequency, or duration, leading to impairments in mental well-being and everyday functioning (Grupe & Nitschke, 2013).

Anxiety is not restricted to clinical diagnoses but exists along a continuum that includes subclinical forms characterised by persistent yet milder symptoms (Volz et al., 2022). Subclinical anxiety is relatively common and remains clinically relevant due to its association with increased risk for developing anxiety disorders (Witlox et al., 2021; Zhong et al., 2024). Throughout this manuscript, the term “anxiety” is used to refer to subclinical anxiety unless otherwise specified, in line with a dimensional framework that differentiates it from clinically diagnosed anxiety disorders.

At the neural level, a substantial body of work has identified key regions implicated in anxiety, including the extended amygdala (including the bed nucleus of the stria terminalis), hippocampus, and ventromedial prefrontal cortex (Holzschneider & Mulert, 2011; Yizhar et al., 2011). However, much of this literature has focused on observable indices such as behavioural and physiological responses. Whether these objective measures correspond closely to the subjective experience of anxiety remains less clear. This poses an important challenge, as defensive behaviours and physiological responses, which are often used as proxies for anxiety do not fully capture conscious emotional experience (Griebel &Holmes, 2013; LeDoux, 2017, 2020; LeDoux & Pine, 2016; Taschereau-Dumouchel et al., 2022)

Here, the two-system framework offers a way to reconcile this discrepancy by proposing that subjective experience and defensive responses are supported by partially distinct, though interacting, neural systems (LeDoux & Pine, 2016). In this account, subcortical and salience-related circuits are primarily involved in threat detection and defensive responding, whereas higher-order cortical regions contribute to the conscious experience of anxiety. Supporting this view, recent task-based fMRI studies using multivoxel pattern analysis have shown dissociable neural representations of subjective and physiological responses (Liu et al., 2024; Taschereau-Dumouchel et al., 2020). However, these studies primarily focused on transient anxiety during the task induced using potent stimuli and have also not examined anxiety-related behavioural responses, such as defensive engagement, which form an important and measurable dimension of anxiety. We adopt the two-system framework's core distinction that subjective experience (i.e., the conscious feeling of anxiety) and defensive and physiological responses to uncertain aversive events are conceptually distinct processes mediated by separable neural circuits. While this framework allows for cross-system interaction rather than complete isolation, we extend it by proposing that the subjective experience of anxiety directly modulates defensive and physiological responses. Specifically, we investigated whether defensive and physiological responses map onto distinct neural circuitry from subjective anxiety after accounting for shared variance between these measures. This approach allowed us to examine whether the neural correlates of defensive and physiological responses remain distinguishable from those associated with subjective anxiety.

Resting-state functional magnetic resonance imaging (rsfMRI) provides a useful approach to address this question by characterising intrinsic functional connectivity across large-scale brain networks. First, unlike task-based fMRI, rsfMRI does not impose additional cognitive or behavioural task demands that may themselves vary as a function of anxiety, vigilance, attentional engagement, or performance strategies (Biswal et al., 1995). This is particularly relevant in anxiety research, where differences in task engagement and anticipatory responding can confound interpretation of task-evoked activation patterns (Grupe & Nitschke, 2013). Second, resting-state paradigms permit the examination of intrinsic large-scale network organisation without requiring explicit task performance or sustained engagement with cognitively demanding tasks inside the scanner (Menon, 2011; Sylvester et al., 2012). Third, resting-state intrinsic connectivity patterns reflect stable and behaviourally meaningful large-scale brain network organisation. These patterns parallel task-evoked activity while also capturing baseline interactions among cognitive, affective, and interoceptive systems, thereby providing complementary insights into the neurobiology of cognitive functions (Smith et al., 2009). Here, our approach of rsfMRI was particularly relevant to examining the heterogeneity across anxiety response systems proposed by the two-system framework, as intrinsic connectivity may reflect enduring network-level organization underlying subjective and objective dimensions of anxiety beyond context-specific task activation.

In the present study, we examined whether behavioural, physiological, and subjective dimensions of subclinical anxiety map onto distinct patterns of intrinsic brain connectivity. Importantly, we specifically examined subclinical anxious affect, operationalised via the National Institutes of Health (NIH) Toolbox Fear-Affect measure, which captures individual variations in affect experienced over brief to intermediate time spans (i.e., days to weeks) rather than stable traits (trait-anxiety) or immediate reactions to prevailing situations (state-anxiety). Here, we consider the Fear-Affect measure as a proxy of the subjective experience of anxiety. The rationale for using this measure is discussed in the Methods section. Using a threat anticipation paradigm, we assessed behavioural responses and physiological arousal alongside self-reported anxiety. Guided by the two-system framework, we tested whether (1) anxiety is associated with behavioural responses under temporally uncertain threat, (2) anxiety is associated with physiological arousal under the same condition, and (3) behavioural, physiological, and subjective components of anxiety correspond to dissociable neural networks. By bringing these dimensions together, the study aimed to better characterise the often-observed discordance across different components of anxiety at the level of intrinsic brain organisation. Here, we show that anxiety is associated with defensive engagement, while no corresponding relationship emerges with physiological arousal. At the neural level, anxiety-modulated objective responses and subjective experience map onto functionally and spatially distinct intrinsic networks, indicating discordance across behavioural, physiological, and experiential dimensions. We interpret these results to more comprehensively characterise the dissociation across the aforementioned dimensions of anxiety at the level of intrinsic connectivity.

## MATERIALS AND METHODS

### Participants

The participants (*n* = 47; 12 females) were recruited based on the following inclusion criteria, (1) age between 18 and 35 years, (2) normal or corrected-to-normal vision with no colour blindness, (3) no reported diagnosis of neurological/psychiatric disorder in the last three months with no history of epilepsy, and (4) not currently taking prescription medications affecting the nervous system and/or respiration (asthma). Behavioural and skin conductance data were collected on Day 1 during a behavioural experiment lasting approximately 40 minutes. The MRI session was conducted on Day 2, with each participant undergoing structural and functional scans totalling about 30 minutes. The interval between the behavioural and MRI sessions averaged 5.2 days, with a standard deviation of 4.3 days. This delay between the behavioural and MRI sessions was primarily due to the absence of an on-campus MRI facility; limited availability of time slots at the off-campus MRI scanning centre, and logistical constraints associated with scheduling both the lengthy in-lab behavioural experiment (~60 minutes) and the MRI session (~30 minutes), including participant travel time to the scanning facility (~ 30 minutes). Three subjects were excluded from the final analysis because of poor signal quality of skin conductance data, and the final sample size reported is after this exclusion [see Table 1 for demographic details].

To determine the required sample size a priori, we identified two prior studies that closely aligned with the design and outcome measures of our study. Please note that, given our study integrates behavioural, physiological, and neuroimaging measures, the effect sizes were drawn from prior studies with closely aligned designs that collectively captured these domains and examined related constructs. Effect sizes from these studies were converted to Pearson's *r* where necessary, as described earlier (Cohen, 1988). First, a neuroimaging study of 39 individuals with trait-anxiety found that the interaction between threat anticipation and anxiety was associated with activation across multiple brain regions, with minimum reported effect size of r = 0.42 (Geng et al., 2018). Second, a neuroimaging study of vigilance in 45 trait-anxious individuals reported correlations between anxiety scores and neural sensitivity to increasing line height across threat-related regions, the basal forebrain ($r$ = 0.50) and prefrontal cortex ($r$ = 0.47) (Somerville et al., 2010).

We powered the study conservatively using the smallest effect size reported from the most comparable prior works above ($r$ = 0.42) to ensure adequate sensitivity. An a priori power analysis was conducted in G*Power version 3.1.9.7 (Faul et al., 2007) for a two-tailed bivariate correlation test with $\alpha$ = 0.05 and statistical power = 0.80. This analysis indicated that a minimum sample size of N = 42 participants was required to detect an effect of this magnitude. The final sample in the present study comprised 47 participants.

This study and its procedures were carried out in accordance with institutional guidelines that received approval from the Institutional Ethics / Review Board of Indraprastha Institute of Information Technology Delhi, India (Approval No. EC/NEW/INST/2024/DL/0440, dated 19 April 2024). Informed written consent was obtained from all participants prior to data collection.

**Subjective Ratings**

We measured the depressive symptoms using the Patient Health Questionnaire (PHQ-9) and excluded individuals with a score greater than 15 (Kroenke et al., 2001). Following that, we employed the Fear Affect scale of the National Institutes of Health (NIH) Toolbox to evaluate the cognitive component of anxiety, assessing self-reported fear and anxious misery. Please note that the Fear Affect scale demonstrates strong convergent validity with established anxiety measures, showing high Pearson correlations with the Generalized Anxiety Disorder-7 (GAD-7; $r = 0.86$) and the Mood-Anxiety State Questionnaire (MASQ; $r = 0.82$), while also exhibiting high internal consistency (Cronbach's $\alpha = 0.95$) (Pilkonis et al., 2013). Importantly, the strong association with the GAD-7 is observed despite differences in symptom assessment windows, with the Fear Affect scale indexing anxiety-related experiences over the past 7 days and the GAD-7 assessing symptoms over the past 14 days. This convergence across partially non-overlapping time windows suggests that the Fear Affect scale captures relatively stable and trait-relevant dimensions of anxiety rather than only transient state fluctuations. The rationale for choosing the Fear Affect scale over the GAD-7 is that the latter primarily assesses the severity and frequency of generalized anxiety symptoms (Spitzer et al., 2006), whereas the Fear Affect scale more directly captures subjective affective experiences of anxiety. The Fear Affect scale was also preferred over the State-Trait Anxiety Inventory (STAI) and Beck Anxiety Inventory (BAI), given concerns regarding the construct specificity of the STAI (Knowles & Olatunji, 2020) and the greater focus of the BAI on somatic anxiety symptoms (Muntingh et al., 2011). The MASQ (Watson & Clark, 2012) was not used because its lengthy assessment would have placed an additional burden on participants and lengthened the testing session; instead, the Fear Affect scale was selected as a shorter and more focused measure that demonstrates strong convergent validity with the MASQ (r = 0.82), while offering greater feasibility for the present study.

In our study here, the interval between Fear Affect assessment and resting-state fMRI acquisition was on average 5.2 days with a standard deviation of 4.3 days, with most participants assessed within a reasonable time window for capturing stable, trait-like anxiety-related individual differences relevant to intrinsic functional connectivity. From a cognitive neuroscience perspective, resting-state network organisation is generally understood to reflect enduring large-scale functional architecture associated with affective and cognitive traits, rather than momentary fluctuations alone. Taken together, these considerations support the interpretation that the Fear Affect scores in the present study primarily reflected trait-like anxiety-related individual differences that were meaningfully associated with the observed resting-state fMRI connectivity patterns. Participants reported on their symptoms on the fixed-form version of the Fear Affect scale on a Google form using a 1-to-5 Likert scale, where 1 indicated 'Never' and 5 indicated 'Always'. Responses to all items were mandatory to ensure complete data and consistency across participants. Fear Affect scale scores were calculated by summing the raw item scores (see Table 1). Higher scores reflected a relatively higher level of anxiety.

**Behavioural experiment**

The behavioural experiment followed the completion of self-report questionnaires. Participants sat at a distance of 57 cm from a 23-inch monitor (1920 × 1080 pixels; 60 Hz; ACER) with their heads stabilised through a chin rest to carry out the behavioural experiment designed on Psychopy (version 2022). The behavioural task (see Figure 1) began with the appearance of a fixation stimulus for 200 ms, followed by a cue X or O for 2 s, indicating the Threat (negative) or Safety (neutral) condition, respectively. An anticipation window appeared after the cue with varying time intervals of 4-6 s during which integers appeared either in a sequential (Certain) or in a random fashion (Uncertain). In the Certain condition, the countdown began from a particular integer and progressed to 1, immediately after which an image (8 × 11° or 11× 8° visual angle) appeared for 500 ms. In the Uncertain condition, integers were presented randomly, with the image appearing unpredictably at any time during the sequence. Sequential presentation in the Certain condition conveyed temporal predictability, whereas random presentation in the Uncertain condition ensured unpredictability (Hur et al., 2020). The participants were instructed to pay attention to the cue and during the anticipation window. They were to respond with a mouse right click in the Safety condition and a mouse left click in the Threat condition as soon as the image appeared on the screen. The trial ended with a participant's response. The next trial began after the inter-trial interval which ranged randomly between 5 to 7 s. Physiological responses were measured concurrently with the behavioural experiment. The analysis was performed during a pseudorandomized 4–6 s anticipation window. A 5 s segment (from 1 s post-cue to 4 s post-anticipation window onset) was analysed, as this represented the shortest common interval across trials. Reaction time was recorded from the onset of the Threat or Safety visual stimulus until the mouse click to quantify behavioural responses. A 1-4 Likert rating scale appeared for 5 s pseudo randomly in eight out of 32 trials per session in a participant to assess aversion to temporal uncertainty during the anticipation window. The scale was included in only a subset of trials to minimize continuous monitoring, which could otherwise attenuate or flatten the affective response associated with uncertainty especially in response to a mild stimulus. To quantify aversiveness to uncertainty, ratings from all four sessions were pooled to calculate the mean rating for each of the four conditions. Lower rating scores indicated lower aversion to uncertainty, whereas higher rating scores indicated greater aversion to uncertainty. As expected, participants on average rated the temporally Uncertain trials to be more aversive than the Certain trials. Particularly, participants rated temporally Uncertain Threat trials (median ± interquartile range or iqr = 1.37 ± 0.86) as more aversive than Certain Threat trials (median ± iqr = 1.25 ± 0.63); Wilcoxon-signed rank test: $Z = 3.70$, $p < .001$. They also rated temporally Uncertain Safety trials (median ± iqr = 1.25 ± 0.75) as more aversive than Certain Safety trials (median ± iqr = 1.13 ± 0.50); Wilcoxon-signed rank test: $Z = 3.28$, $p = .001$, and Uncertain Threat trials as more aversive than Uncertain Safety trials (Wilcoxon-signed rank test: $Z = 2.74$, $p = .006$).

There were eight trials for each of the four conditions (Uncertain Threat, Uncertain Safety, Certain Threat, and Certain Safety) in one session. The total number of sessions was four with intermittent breaks of ~ 4 minutes. The negative (mean valence = 2.42, mean arousal = 6.65) and neutral (mean valence = 5.06, mean arousal = 4.81) images used in the task to indicate the Threat and Safety conditions, respectively, were sourced from International Affective Picture System (IAPS; (Lang et al., 2005)) and were validated in a separate group of 15 participants with demographics comparable to those part of the main study, but were not included in it.

**Behavioural data analysis**

To quantify behavioural response, we computed reaction time of correct trials after pooling data across all four sessions. Participants responded faster in the Certain condition (Certain Threat: 229.4 ± 10.8 ms (mean ± SD); Certain Safety: 233.02 ± 10.62 ms (mean ± SD)) compared to the Uncertain condition (Uncertain Threat: 360.70 ± 6.21 ms (mean ± SD); Uncertain Safety: 362.47 ± 8.09 ms (mean ± SD)) as visible in the raw reaction time. Participants were instructed to attend to the anticipatory period and to respond as quickly as possible once the image appeared. Thus, participants responded to the stimulus (image) only after its onset. A mouse click measured from the offset of the anticipatory period, immediately terminated the presentation of the threat or safety image. Previous research on temporal preparation proposes that temporal predictability enhances multiple stages of information processing (attentional allocation, sensory processing, decision formation, and response preparation) rather than merely allowing participants to pre-execute the motor response (Bausenhart & Ulrich, 2026). Therefore, shorter reaction time under temporal certainty suggests that participants were better prepared to process the visual stimulus when it appeared, rather than because they responded before stimulus onset. The difference in reaction time between the temporal-certainty and temporal-uncertainty conditions, therefore, reflects the intended effect of manipulating temporal predictability. Thus, temporal certainty was associated with more efficient temporal preparation reflected in faster reaction time, whereas temporal uncertainty resulted in slower reaction times, indicating a trend towards less efficient temporal preparation. Subsequently, for each session, the z-scores were computed for correct trials in which a participant achieved at least 27 or more correct responses out of the total 32 trials, accounting for response latencies associated with mouse-based input. The z-scores of all four sessions were pooled to compute the mean reaction time for each condition separately. Lower z-scores indicated faster reaction times. The z-scores pooled across the four sessions were then used for correlational analyses with participants' self-reported anxiety scores to examine whether behavioural responses varied as a function of anxiety. Resulting *p*-values were Bonferroni-corrected for four comparisons.

**Skin conductance data analysis**

Skin conductance was recorded at 100 Hz from Shimmer 3 GSR from the index and middle fingers of the non-dominant hand while the participants performed the behavioural task. After linearly detrending the raw signal, a low bandpass Butterworth filter (implemented with MATLAB's 'butter' function; Lyons, 2004) of first order with a cut-off frequency of 0.05 Hz was applied to smooth the data, and the data was downsampled to 10 Hz as explained elsewhere (Eidlin Levy & Rubinsten, 2021). The tonic component of skin conductance measure represents the slowly varying baseline level of skin conductance, reflecting relatively sustained changes in electrodermal activity. While the phasic component represents transient changes in skin conductance driven by sudomotor nerve activity; in the cvxEDA model, this activity is represented by a sparse, non-negative sudomotor nerve activity (Greco et al., 2016). The tonic and phasic components of the skin conductance signal were segregated using the convex optimisation algorithm implemented with MATLAB's cvxEDA function (Greco et al., 2016). The phasic component of each participant was standardized to remove artifacts and analysed as described elsewhere (Eidlin Levy & Rubinsten, 2021; Grupe & Nitschke, 2011). The correct trials were chosen such that the detectable phasic response was ≥ 0.01 during the anticipation window of 5 s (see Figure 1-E). Then, a baseline correction was performed using the subtraction method. SCRs ≥ 0.01 were extracted from 0 s to 1s from the onset of the cue (X or O), after which the mean baseline SCR was computed.  Next, a cubic spline interpolation was done in the anticipation window of 5 s between the starting and end point of each phasic response (Silva et al., 2023).  Finally, the area under the curve (AUC) was computed by integrating the phasic responses between the starting and end point (Equation 1). In case of more than one phasic measure, cumulative sum of all the AUCs was calculated over the entire anticipation window of 5 s.

$AUC = \int_{t^0}^{t^n} A(t)dt$ (Equation 1)

where, $t^0$= starting point of phasic component;

$t^n$ = end point of phasic component;

$A(t)$ = amplitude at time $t$ within the response window.

A higher AUC reflected high arousal or hypervigilant response to the situation. Three participants were excluded from the analysis due to poor signal quality(Morgan, 2017) as explained in the Participants section. The AUCs calculated across all four sessions were pooled, and then subsequently a mean AUC was computed for each participant and condition. Individuals with detectable responses (≥ 0.01 threshold) on at least 20% of the total number of trials (i.e., 128 trials) were included in the final analysis (n = 47). A sample analysis of AUC calculation of a participant is shown in Supplementary Figure S1, illustrating the above-mentioned procedure. As electrodermal activity is inherently noisy, we compared our data statistically to those from a previous study bearing some similarity with our study design and observed that our data were not noisier (see Supplementary S2 for details of the procedure). Subsequently, the participants' AUCs indicating the arousal level were correlated with their respective anxiety scores obtained through self-report to examine whether physiological responses varied as a function of anxiety.

**Task Validation Across Anticipatory Anxiety Rating (Subjective Aversiveness), Reaction Time, and Skin Conductance**

We first examined whether temporal uncertainty and threat independently or interactively influenced subjective aversiveness, behavioural responses, and physiological arousal. For each measure, we tested the main effects of temporal information (uncertain vs. certain), emotion (threat vs. neutral), and their interaction using 2 × 2 repeated-measures analyses of variance (ANOVAs).

For subjective aversiveness ratings, the residuals deviated from normality (Lilliefor's test; $p = 0.001$). Given the robustness of repeated-measures ANOVA to moderate departures from normality in balanced within-subject designs, we proceeded with the analysis. There were significant main effects of temporal information, $F$ (1, 46) = 10.67, $p = 0.002$, and emotion, $F$ (1, 46) = 11.61, $p = 0.001$. The temporal information × emotion interaction was not significant, F (1, 46) = 0.04, $p = 0.824$ (Figure 2A). Thus, participants rated uncertain trials as more aversive than certain trials and threat trials as more aversive than neutral trials, with no evidence that the effect of threat differed as a function of temporal information.

For reaction time, the residuals did not significantly deviate from normality (Lilliefors test, $p = 0.50$). The 2 × 2 repeated-measures ANOVA revealed a significant main effect of temporal information, $F$ (1, 46) = 610.95, $p < 0.001$. Neither the main effect of emotion, $F$ (1, 46) = 1.85, $p = 0.12$, nor the temporal information × emotion interaction, $F$ (1, 46) = 0.004, $p = 0.95$, was significant (Figure 2B). Participants therefore responded more slowly under temporal uncertainty, irrespective of whether the anticipated stimulus was threatening or neutral.

For phasic skin conductance responses, the residuals also deviated from normality (Lilliefors test, $p = 0.001$), but again, we proceeded with the repeated-measures ANOVA given its robustness to moderate non-normality in balanced within-subject designs. Neither the main effect of temporal information, $F$ (1, 46) = 0.10, $p = 0.922$, nor the main effect of emotion, $F$ (1,

46) = 0.29, $p$ = 0.596, was significant. However, a significant temporal information × emotion interaction was observed, $F$ (1, 46) = 6.76, $p$ = 0.012 (Figure 2C). This interaction indicates that the effect of threat on phasic physiological arousal differed between uncertain and certain conditions, despite the absence of significant main effects of either factor.

**MRI experiment**

Scans were acquired using a 3 Tesla Signa Architect (GE) scanner with a 32-channel head coil. T1-weighted anatomical images were collected with the following parameters: voxel size = 0.5 × 0.5 × 0.5 $mm^3$, 344 slices, repetition time (TR) = 2,500 ms, echo time (TE) = 2.4 ms, flip angle = 25°, and matrix size = 512 × 512. T2-weighted resting state functional scans were obtained over approximately 11 minutes (eyes open) using an echo-planar imaging (EPI) sequence with these parameters: TR = 2,000 ms, TE = 30 ms, flip angle = 90°, resolution matrix = 64 × 64, field of view (FOV) = 200 × 200 $mm^2$, 33 slices, and voxel size = 3.1 × 3.1 × 5 $mm^3$. Resting-state fMRI data were collected in a single session at Mahajan Imaging Labs, SDA, New Delhi, India. We specifically instructed participants to relax and refrain from fixating on any particular thoughts or concerns during the scanning session. During the post-scan debriefing, none of the participants reported experiencing anxiety or distress while in the scanner.

Preprocessing was conducted using the default pipeline in the CONN toolbox (version-2022a; (Whitfield-Gabrieli & Nieto-Castanon, 2012)) implemented in MATLAB (version R2024a). The steps included: (1) functional realignment and unwarping to correct for head motion, (2) slice-timing correction, (3) outlier detection based on framewise displacement exceeding 0.9 mm and global signal deviations ±5 standard deviations from the mean, (4) segmentation and normalization of both functional and structural images, and (5) spatial smoothing of functional scans using an 8 mm full-width at half-maximum Gaussian kernel. To mitigate noise in the first-level voxel-to-voxel independent component analysis (ICA), which is highly susceptible to voxel-level artifacts when deriving subject-specific networks, we applied an 8-mm smoothing kernel. This step agrees with standard preprocessing recommendations for functional neuroimaging data (Alahmadi, 2021; Matsuoka et al., 2021). Following preprocessing, denoising was performed by regressing out confounding effects using the default-pipeline of CONN toolbox. First, physiological noise was modelled using the anatomical component-based noise correction approach. Five noise components were extracted separately from white matter and cerebrospinal fluid (CSF) regions. Second, head-motion effects were modelled using 12 parameters: the six rigid-body motion parameters, comprising three translational parameters (x or left-right, y or anterior-posterior, and z or superior-inferior) and three rotational parameters (pitch, roll, and yaw) and their corresponding first-order temporal derivatives. Third, outlier scans identified during preprocessing were accounted for using scrubbing, with one nuisance regressor assigned to each identified outlier scan. Finally, session-specific constant and linear trends were included to account for baseline differences and slow temporal drifts in the BOLD signal. Temporal band-pass filtering (0.008–0.09 Hz) was then applied to remove low- and high-frequency noise as part of the denoising procedure. Participants' head motion was minimal, with a mean framewise displacement (FD) of 0.12 ± 0.04 mm. The group-level Independent Component Analysis (ICA) was conducted by setting the number of ICA factors to 40 and dimensionality reduction parameter to 64 during first-level analysis. The ICA networks were identified using the 'Summary' option available in the second-level results. Subsequently, the spatial correlations between each ICA component and the default CONN network templates were calculated using the "Compute Spatial Match to Template" function available in the ICA tools. Five ICA components (out of 40) corresponding to the best match for default mode network (DMN), salience network (SN), frontoparietal network (FPN), cerebellar network (CN), and visual network (VN) were identified using the CONN toolbox's spatial match-to-template procedure and the 'best match' to the default CONN network template was selected, followed by visual inspection of the resulting spatial maps. For each network, the ICA component showing the highest spatial correlation (measured using the dice coefficient) with the respective default CONN network template was considered the best match. Visual inspection was then performed for each subject to confirm that the spatial distribution of the component was consistent with the canonical anatomical distribution of the target network, based on its characteristic cortical and/or subcortical regions. The selected ICA networks were entered into a second-level general linear model (GLM) to evaluate their associations with the independent variable(s) for the group-level effects.

**MRI statistical analyses**

To identify brain networks, we first conducted a systematic search on PubMed using the terms (“anxiety”) AND (“functional magnetic resonance imaging” OR “fMRI”) AND (“two-system framework”), restricted to studies published between 2000 and 2026. This search did not yield studies that directly combined the two-system framework with network-based functional neuroimaging analyses of anxiety, rather than indicating an absence of anxiety-related fMRI studies more generally. We therefore adopted the approach of theory-driven targeted review of key conceptual and empirical papers along with their cross-referenced literature to define the brain networks. Specifically, we first identified the foundational conceptual paper explaining the two-system framework of anxiety (LeDoux & Pine, 2016) and subsequent empirical studies explicitly examining dissociations between subjective and objective dimensions of anxiety or fear using neuroimaging approaches (Taschereau-Dumouchel et al., 2020, 2022). We then examined the references and cross-citations within these papers to identify large-scale networks that were consistently implicated across studies in threat detection, defensive responding, interoception, and subjective emotional experience. Guided by the two-system framework, we identified five large-scale functional networks corresponding to the systems governing defensive (objective) responses and subjective experience of subclinical anxiety: (a) salience network; (b) default mode network; (c) frontoparietal network; (d) visual network; and (e) cerebellar network. These networks were selected because they collectively capture the neural processes underlying the development, persistence, and clinical manifestation of anxiety and have been reported to show altered functional connectivity in individuals with subclinical anxiety symptoms. Anxiety is now understood as a systems-level disorder involving alterations in large-scale functional brain networks responsible for threat detection, self-referential cognition (autobiographical memory), executive control (biased attention to threatening stimuli), and sensory processing. The rationale for including each network is outlined as follows: (a) salience network is linked to detection of biologically relevant or threatening stimuli, autonomic arousal, and interoceptive awareness, thereby has a role in defensive responses and physiological symptoms of anxiety (Sylvester et al., 2012; Xia et al., 2017); (b) default mode network contributes to self-referential processing, autobiographical memory, and repetitive negative thinking, processes that maintain anxious states through persistent anticipation of future threat (Dixon et al., 2022; Sylvester et al., 2012; Tao et al., 2015); (c) frontoparietal network has a role in executive control and top-down regulation of emotional responses; dysfunction within this network impedes cognitive control thereby impairing adaptive emotion regulation and aggravating anxiety (Dixon et al., 2022; Sylvester et al., 2012); (d) visual network's involvement in early perceptual processing and attentional bias toward threatening stimuli influence cognitive appraisal as a result of interaction between sensory information and experience (Taschereau-Dumouchel et al., 2024); (e) cerebellar network has a role beyond motor coordination to support modulation of cognitive-affective processes through interaction with key regions involved in cognitive and emotional processing, contributing to behavioural and experiential characteristics linked to anxiety (Dhar et al., 2024; Mir et al., 2026; Phillips et al., 2015). Together, these networks encompass complementary mechanisms involved in threat perception, emotional regulation, cognitive maintenance, and behavioural and physiological expression, providing a comprehensive framework for investigating anxiety-related neurobiology. These five networks were identified initially based on the two-system framework and their contribution to defensive and physiological responses and subjective anxiety-related experience. To substantiate their relevance to anxiety, we subsequently conducted a targeted literature search on PubMed within the time range between 2000-2026 using the search terms detailed in Table 2. This search served as a verification step to assess whether each of the theoretically identified networks has previously been implicated in anxiety-related processes or symptoms. The literature search was carried out to establish an empirical basis for investigating the networks identified from the conceptual framework.

The multiple regression models were built controlling for age, gender and delay between behavioural and MRI experiment (in number of days) as nuisance covariates. First, we assessed whether anxiety moderated the association between behavioural response and rsFC by including an anxiety x behavioural response interaction term (Equation 2). Second, we investigated whether anxiety moderated the association between physiological response and rsFC by including an anxiety x physiological response interaction term (Equation 3). Third, we tested the association between anxiety and rsFC while controlling for the modulatory effects of anxiety both on the behavioural and physiological responses (Equation 4).

$$y = \beta 0 + \beta 1\,(age) + \beta 2\,(gender) + \beta 3\,(delay) + \beta 4\,(anxiety) + \beta 5\,(behavioural\,response) + \beta 6\,(anxiety \times behavioural\,response) + \epsilon$$

(Equation 2)

$$y = \beta 0 + \beta 1\,(age) + \beta 2\,(gender) + \beta 3\,(delay) + \beta 4\,(anxiety) + \beta 5\,(physiological\,response) + \beta 6\,(anxiety \times physiological\,response) + \epsilon$$

(Equation 3)

$$y = \beta 0 + \beta 1\,(age) + \beta 2\,(gender) + \beta 3\,(delay) + \beta 4\,(anxiety) + \beta 7\,(anxiety \times behavioural\ response) + \beta 8\,(anxiety \times physiological\,response) + \epsilon$$

(Equation 4)

where y = rsFC between parent network and cluster; β0 = intercept; β1-β8 = parameter estimates, and ϵ = residual. Age, gender, and delay between behavioural and MRI experiment were covariates of no interest in all regression models (Equation 2, 3, and 4). The first regression model (Equation 2) measured how the interaction between behavioural response to anticipation of aversive event and anxiety associated with rsFC after controlling for the main effects of behavioural response and anxiety. The second regression model (Equation 3) measured how the interaction between physiological response to anticipation of aversive event and anxiety associated with rsFC after controlling for the main effects of physiological response and anxiety. The third regression model (Equation 4) measured the influence of anxiety i.e., the subjective experience, after controlling for the modulatory effects of anxiety on objective responses. The independent variables in the model showed no evidence of multicollinearity, as variance inflation factors (VIFs) were less than 4 (all VIFs < 1.12). Please note that we have only reported the MRI results by correlating the behavioural / physiological measures in response to the Uncertain Threat condition as anxiety originates from anticipation about a potential threatening event. The rationale behind focusing exclusively on intrinsic networks associated with the Uncertain Threat condition is theoretically grounded in the distinction between anticipatory anxiety and immediate threat processing (i.e., Certain Threat). As explained earlier, anxiety is primarily characterized by heightened sensitivity or reactivity to uncertain, unpredictable, and temporally diffuse threats, rather than to clearly defined threat (Hur et al., 2020; LeDoux & Pine, 2016). The Uncertain Threat condition specifically dealt with this anticipatory state, as it involved the expectation of a potential aversive event without precise information about its timing. In contrast, the Certain Threat condition measured responses to predictable and imminent danger, which are more closely aligned with acute fear processing rather than anxiety. Similarly, the Uncertain Safety condition associated with ambiguity in the absence of threat, which may not reliably elicit anticipatory anxiety, and the Certain Safety condition represents a baseline state with minimal threat-related processing. By isolating the Uncertain Threat condition and correlating MRI-derived intrinsic networks with behavioural and physiological measures, the present study investigated neural systems specifically involved in anticipatory anxiety. This approach allows a more precise characterization of the networks underlying anxiety-related anticipation. Additionally, the significant results observed for the behavioural analysis provide evidence that the responses vary particularly for the Uncertain Threat condition as a function of anxiety.

To evaluate rsFC, a Group Independent Component Analysis (ICA) procedure was implemented within the CONN toolbox to identify distinct parent networks and characterise their network-to-voxel connectivity profiles. To account for the multiplicity of testing within our independent hypotheses, a stepwise multiple-comparisons correction strategy was implemented. To control for the false positive rate within each isolated network map, spatially, individual second-level parametric maps for each ICA-derived network were thresholded using a voxel-level height threshold of $p < 0.001$ (uncorrected) and a cluster-level extent threshold of $p$-FDR < 0.05. To further control for the false discovery rate while running five separate network models under an independent hypothesis, we submitted the $p$-values from all tested parent networks into an additional Benjamini-Hochberg False Discovery Rate (FDR) procedure implemented using the MATLAB function 'mafdr' (Benjamini & Hochberg, 1995). For networks that yielded no clusters surviving the initial spatial thresholding criteria, a conservative $p$-value of 1.0 was assigned. This step ensures that networks with completely null spatial maps still penalise the overall threshold, preventing the artificial inflation of statistical significance. Following this final correction step, the unadjusted cluster-level $p$-values and the final adjusted hypothesis-wide $p$--FDR values (across all five models) were reported in Table 3.

## RESULTS

### Behavioural results: Anxiety-Behaviour associations in Threat and Safety conditions

To measure the influence of anxiety on behaviour, we correlated anxiety measured using a self-report scale with reaction time measured through the behavioural task. After confirming normality for anxiety scores and behavioural responses (Lilliefor's test; all $k$s < 0.12; all $p$s > 0.07) and ensuring there were no univariate outliers (all values within ±3 SD), Pearson's correlation was computed. A negative correlation (Pearson's $r$= -0.47, degrees of freedom or $df$ = 45, 95% confidence interval ($CI$) = [-0.67, -0.21], slope (β) = -0.01, $p$ = 0.001) was observed for the Uncertain Threat condition, indicating that with increasing anxiety, participants responded faster in anticipation of temporally uncertain aversive events, and this association remained significant after applying Bonferroni-correction ($p$ = 0.05 / 4 ≈ 0.013). This result suggests that increased defensive engagement observed during the preceding uncertainty period associated with an aversive event may have contributed to a defensive, escape-like behaviour in individuals with elevated anxiety. As the mouse click terminated presentation of the unpleasant image, the faster responses observed in highly anxious participants may reflect a greater tendency to escape or terminate prolonged exposure to the aversive stimulus. No correlation was observed between anxiety scores and reaction time of Uncertain Safety (Pearson's $r$ = -0.21, $df$ = 45, 95% $CI$ = [-0.47, 0.09], slope (β) = -0.01, $p$ = 0.16), Certain Threat (Pearson's $r$ = 0.20, $df$ = 45, 95% $CI$ = [-0.09, 0.46], slope (β) = 0.01, $p$ = 0.20), and Certain Safety (Pearson's $r$ = 0.10, $df$ = 45, 95% $CI$ = [-0.20, 0.37], slope (β) = 0.003, $p$ = 0.52) conditions. This pattern suggests that predictable threats may elicit relatively uniform responses across individuals due to shared survival mechanisms.

**Skin conductance results: Anxiety-phasic response associations in Threat and Safety conditions**

To measure correlation between anxiety scores and phasic response, we computed Spearman's *rho*, as the assumptions of normality were violated (Lilliefor's test; all *k*s > 0.14; three *p*s < 0.05). No correlation was observed between the anxiety scores and phasic response associated with Uncertain Threat condition (*rho* = -0.04, *df* = 45, 95% *CI* = [-0.32, 0.25], *p* = 0.81, see Supplementary Figure S2). A plausible explanation for this result could be that the stimulus used may not have been sufficiently aversive to engage the neural circuits responsible for eliciting physiological responses. Prior work has shown that anticipation of a physically threatening stimuli, such as shock or carbon dioxide inhalation, has resulted in exaggerated arousal level in individuals with high-trait anxiety and activation in insula associated with interoception (Somerville et al., 2010; Telch et al., 2011). Further, the correlation observed between anxiety scores and phasic response of Uncertain Safety (*rho* = 0.22, *df* = 45, 95% *CI* = [-0.07, 0.48], *p* = 0.13), Certain Threat (*rho* = -0.007, *df* = 45, 95% *CI* = [-0.29, 0.28], *p* = 0.97), and Certain Safety (*rho* = 0.01, *df* = 45, 95% *CI* = [-0.28, 0.30], *p* = 0.95), conditions were also statistically insignificant.

**MRI results**

***Anxiety moderated association between behavioural response and rsFC***

We examined whether anxiety moderated the association between behavioural response and rsFC to determine whether the relationship between reaction time and rsFC varied as a function of anxiety levels. The moderation analysis revealed a significant positive association between the anxiety-modulated behavioural response and rsFC (Figure 4A; Table 3), indicating that the relationship between reaction time and rsFC became increasingly positive with higher levels of anxiety. The significant rsFC was observed between the default mode network (DMN) and two clusters: one encompassing the bilateral paracingulate gyrus and anterior cingulate gyrus, and another involving the right occipital fusiform gyrus and the left occipital pole within the visual cortex.

To further examine whether this association was specific to uncertain threat, we conducted an additional analysis in which the variance shared with behavioural responses to certain threat and uncertain safety was removed from the uncertain-threat condition through subtraction. The association between anxiety-modulated behavioural responses and DMN–rsFC remained significant, indicating that the observed effect was not attributable solely to variance shared with certain threat or uncertain safety. The results of this analysis are provided in Supplementary Section S7.

### *Anxiety moderated association between physiological response and rsFC*

We examined whether anxiety moderated the association between physiological response and rsFC to determine whether the relationship between arousal level, measured through skin conductance, and rsFC varied as a function of anxiety levels. The moderation analysis revealed a significant positive association between the anxiety-modulated physiological response and rsFC (Figure 4B; Table 3), indicating that the relationship between physiological arousal and rsFC became increasingly positive with higher levels of anxiety. Significant rsFC was observed between the salience network (SN) and a cluster within the visual cortex encompassing the left occipital pole, left superior division of the lateral occipital cortex, left intracalcarine cortex, left cuneal cortex, and left supracalcarine cortex.

Additionally, to examine whether this association was specific to uncertain threat, we conducted analysis in which the variance shared with physiological responses to certain threat and uncertain safety behavioural was removed from the uncertain-threat condition by subtraction. This analysis revealed no significant networks, suggesting that the observed effect was not specific to uncertain threat and may instead reflect broader uncertainty- or threat-related processes.

### *Association between subjective experience of anxiety and rsFC*

We examined the association between subjective experience of anxiety and rsFC to determine whether subjective experience of anxiety was uniquely associated with intrinsic connectivity after accounting for the modulatory effects of anxiety on behavioural and physiological responses. The analysis revealed a significant negative association between subjective anxiety and rsFC between the cerebellar network and two brain clusters. The first cluster encompassed the left superior division of the lateral occipital cortex, left precuneus, left angular gyrus, and left superior parietal lobule, while the second cluster was located in the left frontal pole. Additionally, a significant positive correlation was observed between the cerebellar network and supplementary motor cortex and paracingulate gyrus (Figure 4C;).

We performed an additional analysis to measure the effect of subjective experience of anxiety while controlling for the anxiety-modulated behavioural and physiological responses specific to uncertain threat. The analysis revealed no significant network.

Moreover, we also conducted an additional analysis, which was not part of our primary aims, to examine whether anxiety moderated the association between aversiveness to uncertainty and rsFC. This analysis revealed a positive association between the Default Mode Network and regions of the visual cortex. The details of this analysis are provided in Section S8 of the Supplementary Materials.

**Intrinsic Functional Network Dissociation Analysis**

To evaluate whether the three Independent Component Analysis (ICA) parent networks exhibited distinct functional connectivity profiles, we performed an analysis using the General Linear Model (GLM) Calculator tool within the CONN functional connectivity toolbox (2022.v2047), through the following steps: (a) The spatial shapes of ICA-derived parent Network 1 (Default Mode Network), Network 2 (Salience Network), and Network 3 (Cerebellar Network) were initially exported from the second-level Results Explorer as individual, binarized spatial masks. These spatial masks were then treated as regions of interest (ROI) and the average blood-oxygen-level-dependent (BOLD) time-series from each target area (cluster) was extracted for every participant; (b) A first-level ROI (network) -to-ROI (target) analysis was run to compute the functional connectivity between all three source networks and all three target clusters. For each of the 47 participants, this yielded a 3 × 3 connectivity matrix consisting of 9 distinct path values (b-coefficients representing Fisher-transformed correlation values of, i) *Network 1 paths: Network1toCluster1, Network1toCluster2, Network1toCluster3*; ii) *Network 2 paths: Network2toCluster1, Network2 toCluster2, Network2toCluster3*; iii) *Network 3 paths: Network3toCluster1, Network3toCluster2, Network3toCluster3*); c) At the second level, an analysis was conducted within the CONN Calculator tool to determine if the functional connectivity profiles of the three networks were statistically dissociable. All nine first-level connectivity values (3 x 3 network-by-cluster pathway coefficients) were entered into a single GLM. To evaluate these pathways across the entire cohort, the second-level subject contrast was set to evaluate the group main effect. A cross-interaction contrast matrix was then applied to the nine connectivity values to test for a network-by-cluster interaction effect. This model revealed a significant network-by-cluster interaction, ($F_{(4,43)} = 14.15$; $p < 0.001$) indicating a functional dissociation amongst the networks; d) To identify the specific path differences driving this primary interaction, planned post-hoc repeated-measures contrasts were performed in CONN for each pair of networks (Network 1 vs. Network 2, Network 2 vs. Network 3, and Network 1 vs. Network 3). Each contrast tested whether the two networks exhibited different connectivity paths across the three target clusters. Family-wise error rate was controlled using Bonferroni method (Dice, 1945; Eichstaedt et al., 2013). These analyses revealed significant differences in connectivity profiles between Network 1 and Network 2 ($F_{(2,45)}$ = 6.99, $p$ = 0.002; Bonferroni adjusted $p$ = 0.008), between Network 2 and Network 3 ($F_{(2,45)}$ = 28.38, $p$ < 0.001; Bonferroni adjusted $p$ = <0.003), and between Network 1 and Network 3 ($F_{(2,45)}$ = 9.09, $p$ = 0.001; Bonferroni adjusted $p$ = 0.003), indicating that all three networks exhibit distinct functional connectivity paths across the target regions. The reported Bonferroni-adjusted *p*-values account for the three pairwise comparisons conducted.

The significant omnibus interaction and pairwise tests establish an intrinsic, within-subject network dissociation. Within the functional architecture of the brain by cross-evaluating three active network nodes against three independent spatial target clusters within the individual, we demonstrate that changing the source network systematically alters the connectivity strength to the target clusters. This interaction rules out the possibility that the three networks are minor variations of a single macro-system, instead confirming three distinct functional boundaries as explained below.

First, because the analysis explicitly compares the slopes between different functional pathways, the significant interaction proves that these connectivity profiles are not parallel. Second, this analysis mitigates spatial overlap bias which is a limitation in ICA-based fMRI where proximal regions share artificial correlation variance due to spatial smoothing. By explicitly mapping the time-series of each network onto the spatial territory of alternative, mismatched clusters (cross-over nodes), the model isolates variance unique to each specific network-to-target connectivity. The significant interaction proves that the functional coupling is driven by localised communication streams rather than artificial spatial overlap. Also, by testing all nine functional communication pathways the analysis confirms that no single network shares a duplicate connectivity profile with another. Therefore, the three ICA-derived parent networks exhibit distinct, non-parallel functional connectivity profiles across their respective target clusters, establishing a statistically significant functional network dissociation.

## Spatial Dissociation Analysis

To assess whether the identified functional connectivity pathways represented distinct large-scale brain systems, we quantified the spatial overlap between the parent network–target cluster pairs. For each significant connectivity pathway, the parent network and its associated target cluster were combined to form a single macrosystem. Pairwise spatial overlap between macrosystems was quantified using the Sorensen-Dice coefficient (Dice, 1945), computed with MATLAB's built-in ‘dice’ function. The Dice coefficient measures the degree of spatial similarity between two binary images by calculating twice the number of overlapping voxels divided by the total number of voxels across both images. The resulting Dice coefficients were 0.04, 0.004, and 0.006 for Macrosystems A-B, B-C, and A-C, respectively, indicating negligible spatial overlap and demonstrating that the three macrosystems exhibited largely distinct spatial topographies.

To further quantify their anatomical separation, we computed the Euclidean distance between the intensity-weighted centres of mass of each macrosystem using MATLAB's ‘norm’ function (Heiberg & Fitzpatrick, 2008). For each thresholded spatial map, all voxels with positive values were included, and the centroid was calculated as the intensity-weighted average of voxel coordinates, such that voxels with higher intensity contributed proportionally more to the centroid location. The centroid coordinates were subsequently transformed from voxel space to MNI millimetre space using the affine transformation matrix, after which pairwise Euclidean distances between macrosystem centroids were calculated. The distances between the centroids of Macrosystems A and B, B and C, and A and C were 66.36 mm, 77.89 mm, and 35.10 mm, respectively. Together, these findings indicate that the identified macrosystems are not only functionally dissociable but also anatomically segregated, occupying distinct spatial locations within the brain.

## DISCUSSION

The present study sought to determine whether the behavioural, physiological, and subjective dimensions of anxiety are supported by common or dissociable intrinsic functional brain networks in a subclinical population. To address this question, we combined an ecologically relevant anticipation paradigm based on temporal uncertainty of mildly aversive visual stimuli with resting-state functional connectivity analyses and subjective anxiety ratings. We assessed anxiety-modulated objective responses across two levels – behavioural indices and autonomic reactivity marked by skin conductance responses by employing an ecologically relevant design based on the temporal uncertainty of relatively mild aversive visual stimuli rather than an explicit threat paradigm, along with the subjective experience of anxiety measured through a self-report questionnaire. Building on prior work demonstrating dissociable neural networks for subjective experience and objective responses in task-based fMRI (Liu et al., 2024; Taschereau-Dumouchel et al., 2020), we first examined the relationship between objective (behavioural response and physiological arousal) indices of anxiety and subjective measure of anxiety in a subclinical sample of young adults. We then independently related each of these three dimensions (anxiety modulated behavioural response, anxiety modulated physiological arousal, and subjective distress) to rsFC.

We found that elevated anxiety levels were linked to heightened behavioural defensive engagement during anticipation of Uncertain Threat, whereas no association was observed with physiological arousal. At the neural level, we identified three distinct network-to-voxel rsfMRI connectivity patterns that robustly associate with anxiety-modulated behavioural responses, anxiety-modulated physiological arousal level, and the subjective experience shaped by anxiety, respectively. Importantly, complementary functional and spatial dissociation analyses demonstrated that these connectivity patterns constituted distinct and separable macrosystems, exhibiting negligible spatial overlap and substantial anatomical segregation. Specifically, anxiety-modulated behavioural responses were linked to increased connectivity between the default mode network and (para) cingulate gyrus and regions within the visual cortex. In contrast, anxiety-related physiological arousal was associated with greater connectivity between the salience network and regions within visual cortex and cuneus. Finally, the subjective experience of anxiety was linked to decreased connectivity between cerebellar network and regions within visual cortex, parietal cortex, and frontal cortex. Together, these neurobiological findings may reflect the dissociation between behavioural, physiological, and subjective responses of anxiety (Klumbies et al., 2014; Lang, 1968; Takemura et al., 2026) in subclinical young adults.

A notable feature of the present study is the inclusion of a subclinical population from a non-Western cultural context, which is relatively underrepresented in affective cognitive neuroscience research. As social and cultural factors can shape how individuals experience, express, and respond to threat and uncertainty, characterising the neural networks associated with the behavioural, physiological, and subjective dimensions of anxiety in this population extends the evidence base beyond predominantly Western samples. These findings may therefore provide a useful reference point for future studies examining anxious affect and its neural correlates in similar sociocultural contexts.

However, these findings were primarily observed using the composite uncertain-threat metric that captures multiple latent components of uncertainty, threat-related processing. When a more stringent measure of pure uncertain threat was derived by removing variance shared with certain threat and uncertain safety through subtraction, only the association between anxiety-modulated behavioural responses and DMN-paracingulate gyrus connectivity remained significant. The absence of corresponding effects for physiological arousal and subjective anxiety suggests that these neural correlates may reflect broader uncertainty or threat-related processes rather than processes uniquely attributable to uncertain threat. Moreover, the findings need to be interpreted cautiously, as we could not rule out the possibility that these networks may also reflect participants' experiences of worry during scanning, as we did not collect objective measures, such as eye tracking, to assess participants' behaviour during the scan, nor did we obtain measures of the worry they experienced before or after scanning.

***Associations between subjective experience of anxiety and behavioural and physiological responses***

The observed association between anxiety and behavioural responses during anticipation of Uncertain Threat may indicate increased defensive engagement toward potential aversive events, rather than reflecting intolerance of uncertainty alone. This finding aligns with previous work which has observed similar monitoring of the environment in anxiety (Riedel et al., 2016), particularly under conditions of uncertain threat anticipation (Hur et al., 2020; Somerville et al., 2010). Importantly, participants also rated the temporally Uncertain conditions as more aversive than the Certain conditions, suggesting that the paradigm was successful in eliciting subjective threat anticipation despite the relatively mild nature of the stimuli. Moreover, the significant association observed specifically between anxiety and behavioural response during Uncertain Threat, but not in the other conditions, further supports the interpretation that the paradigm engaged anxiety-relevant anticipatory processes rather than reflecting a generalised response tendency. Despite these convergent behavioural and subjective indicators of anticipatory threat processing, we did not observe a corresponding association between anxiety and physiological arousal.

One possible explanation is that the aversive visual stimuli employed in our study may have preferentially engaged cognitive-attentional aspects of anticipatory anxiety without producing sufficiently intense autonomic activation. Physiological responses are strongly influenced by stimulus intensity and are often more robustly elicited by physically threatening or highly salient aversive stimuli, such as shock anticipation or carbon dioxide inhalation (Somerville et al., 2010; Telch et al., 2011). Additionally, our sample comprised healthy young adults with subclinical rather than clinically diagnosed anxiety and therefore may not have exhibited the heightened autonomic responsivity often observed in clinical anxiety populations. In this context, relatively mild aversive stimuli may have been sufficient to engage anticipatory vigilance and subjective threat appraisal, while remaining below the threshold required to elicit pronounced sympathetic activity. It is worth noting that the absence of a significant physiological activity-anxiety association does not necessarily imply absence of anxiety-related processing, as prior work has documented discordance between behavioural, physiological, and subjective indices of anxiety and fear (Klumbies et al., 2014; Takemura et al., 2026; Taschereau-Dumouchel et al., 2020). Within the framework of the present study, the dissociation between heightened defensive engagement and absent autonomic amplification may therefore reflect the possibility that temporal unpredictability enhanced anticipatory monitoring and cognitive vigilance without proportionally engaging sympathetic arousal systems for relatively mild stimuli.

***Dissociating rsFC correlates of objective (behavioural and physiological) responses modulated by anxiety and subjective experience of anxiety***

Associations observed between anxiety-related dimensions and rsFC suggest that objective defensive responses of anxiety to temporally uncertain aversive visual stimuli and subjective experience of anxiety map onto distinct neural systems. We identified statistically robust associations in three functional connections. First, we observed that anxiety moderated the association between behavioural vigilance and rsFC between the default mode network and clusters encompassing (para) cingulate gyrus and visual cortex. Prior evidence indicates that paracingulate gyrus belonging to limbic system has a role in cognitive and affective regulation, and sensory processing. The increased connectivity between DMN and (para) cingulate gyrus and visual cortex with higher levels of anxiety may suggest an increased alertness to the external environment potentially driven by an increased tendency to monitor for threatening stimuli or a state of anticipatory preparedness based on previous aversive experiences or memories (Azarias et al., 2025; Chadick & Gazzaley, 2011; Wysiadecki et al., 2021). Second, anxiety moderated the association between physiological arousal and rsFC, such that higher anxiety was linked to increased connectivity between the salience network, implicated in emotion and anxiety, and regions within the visual cortex, including the cuneus, as physiological arousal increased. Given the salience network's role in detecting and integrating behaviourally relevant internal and external signals and interoception, this pattern may reflect enhanced allocation of attentional resources to or through visual information during heightened physiological arousal. Such enhanced visual processing might be linked to increased monitoring of bodily states (interoception) and promote attentional reorienting toward potentially relevant or threatening environmental cues (Menon & Uddin, 2010; Seeley, 2019). Lastly, subjective experience of anxiety obtained through self-report was associated with reduced connectivity between cerebellar network and regions within visual, parietal cortex, and frontal cortices alongside increased connectivity with the supplementary cortex, superior frontal gyrus, and regions within limbic system. The connectivity observed between the cerebellar network and higher order cortical regions is noteworthy as the higher-order cortical regions especially frontal pole and parietal lobule have been associated with abstract evaluation and integration of cognitive states. This pattern may reflect altered integration of information between cerebellar and cortical systems involved in evaluating internal cognitive and affective states, potentially contributing to the subjective experience of anxiety. Consistent with this interpretation, recent work from our lab demonstrates that structural variation in cerebellar lobule VI is associated with individual differences in how subclinical anxiety modulate deployment of visuospatial attention, further supporting a role of cerebellar–cortical circuitry in the cognitive-affective modulation (Mir et al., 2026). Conversely, the increased connectivity with limbic regions may indicate greater coupling between cerebellar circuits and affective processing networks linked with heightened emotional processing in individuals reporting higher anxiety. Together, these findings align with the notion of response-system discordance in anxiety, with different components associated with distinct network interactions; however further work is needed to confirm and better characterise these associations.

Although the above-mentioned findings highlight large-scale networks linked to defensive responses modulated by anxiety and the subjective experience of anxiety, their interpretation need to be considered with caution given the sample's gender imbalance (male: female ≈ 3:1; see Table 1), especially in light of documented sex differences in anxiety and emotional processing. Previous studies indicate distinct reactions to emotions and neurobiological profiles in females, such as differences in intrinsic connectivity within regions such as the cingulate cortex, hippocampus and insula, which are part of DMN and salience networks (Sacher et al., 2013; Stevens & Hamann, 2012). In this context, the functional networks identified in the present study, including anxiety-modulated interactions overlap with systems implicated in affective processing. Since we included gender as a nuisance covariate in the rsFC analyses, this limits us to draw inferences regarding sex-specific neural mechanisms. Therefore, the above-mentioned findings need to be interpreted as reflecting a general trend within the sampled population, with the caveat that the strength or direction of these network-level associations may vary in gender-balanced samples.

Another point to consider is that trait impulsivity has previously been associated with heightened physiological arousal to salient or uncertain events and may therefore influence both behavioural and physiological responses (Baker et al., 2024; Zhang et al., 2015). Consequently, trait impulsivity may have acted as a confounding factor in the observed association between anxiety scores and behavioural responses. Future studies may consider controlling for trait impulsivity to better disentangle the contribution of anxiety from that of impulsivity.

***Limitations and Future directions***

Finally, we identify a few limitations in our study which can be addressed in future work. First, our sample had a disproportionately higher number of males, limiting the generalisability of the findings and underscoring the need for replication in a cohort with a more balanced gender distribution. Second, time of day was not included as a nuisance covariate despite evidence for circadian rhythms' influence on rsFC. Third, the resting-state fMRI scans were acquired in a single session due to resource constraints; future work incorporating repeated measurements across multiple sessions would allow assessment of the stability and reliability of the observed connectivity patterns. Fourth, the modest sample size necessitates replication in larger samples to confirm the robustness of the findings. Fifth, future studies could incorporate objective measures, such as eye tracking, alongside pre- and post-scan assessments of subjective worry to determine whether worry or anxiety experienced during scanning may have influenced the observed brain connectivity patterns. Sixth, the hypothesis-driven selection of networks may introduce bias by limiting analyses to predefined networks; future studies using whole-brain approaches and larger samples are warranted to validate these results. Seventh, multiple comparisons remain a concern and may increase the likelihood of false-positive findings, despite applied corrections, which future studies may seek to mitigate by increasing sample sizes, allowing stringent corrections and prioritising independent replication of our results. Eighth, future studies could use more potent aversive stimuli in individuals with clinical anxiety. Lastly, as the findings are correlational, establishing causality and translational potential would require perturbing the identified brain networks using non-invasive stimulation or pharmacological manipulation.

### *Conclusion*

Despite the above limitations, the present study aimed to provide a neurobiological explanation for the response-discordance observed across the three dimensions of subclinical anxiety – behavioural, physiological, and subjective distress using the resting state functional brain networks in a young adult sample. The intrinsic connectivity found suggests dissociable yet overlapping networks underlie the objective responses and subjective experience of anxiety, thereby complementing the limited existing task-based fMRI studies. Future research may examine whether these identified networks facilitate the integration or drive the discordance among the behavioural, physiological, and subjective components of anxiety.

## DECLARATIONS

**Funding Information**

The research was funded by the intramural research Professional Development Allowance of IIIT-Delhi and the Science and Engineering Research Board Core Research Grant to Dr. Mrinmoy Chakrabarty (SERB-CRG/2022/008119). Ms. Shruti Kinger was supported by the institute PhD fellowship by IIIT-Delhi. Ms. Naviya Lall was supported by the research grant (DST/CSRI/2021/201) awarded to Dr. Mrinmoy Chakrabarty. The funders were not involved in the study design, data collection, analysis, interpretation, the writing of this article, or the decision to submit it for publication.

**Competing interests**

The authors declare no competing interests.

**Ethics approval**

This study and its procedures were carried out in accordance with institutional guidelines that received approval from the Institutional Ethics / Review Board of Indraprastha Institute of Information Technology Delhi, India (Approval No. EC/NEW/INST/2024/DL/0440, dated 19 April 2024).

**Consent to participate**

Informed written consent was obtained from all participants prior to data collection.

**Consent for publication**

Participants provided informed written consent for the publication of their data subject to anonymization.

**Data and code availability statement**

The behavioural data and codes that led to the present results can be found here: https://github.com/shrutikinger/Resting-State-Networks-Underlying-the-Experience-and-Expression-of-Subclinical-Anxiety .

The rsfMRI data linked with the present results can be found here https://zenodo.org/records/19666693 .

**Authors' contributions**

**Shruti Kinger:** Conceptualization, Data Curation, Formal Analysis, Investigation, Methodology, Software, Visualization, Writing - Original Draft Preparation

**Naviya Lall:** Data Curation, Investigation, Software, Visualization, Writing - Original Draft Preparation

**Mrinmoy Chakrabarty:** Conceptualization, Formal Analysis, Funding Acquisition, Investigation, Methodology, Project Administration, Software, Supervision, Validation, Visualization, Writing - Review & Editing

**Acknowledgments**

The authors thank all participants for their time and participation in the experiment. The authors are grateful to Manasi Chaturvedi (University of Texas, Austin, USA) and Dr. Suhail Rafiq Mir (All India Institute of Medical Sciences Delhi, INDIA) for their assistance and cooperation during data collection. The authors also thank Dr. Sumitash Jana (Institute of Information Technology, Delhi) for his valuable comments regarding potential noise in the skin conductance data; subsequent validation using an independent dataset did not indicate elevated noise levels. The authors also acknowledge the Centre for Advanced Research in Imaging, Neuroscience and Genomics (CARING) of Mahajan Imaging, SDA, New Delhi, Delhi 110016, for providing the MRI imaging facility and Ms. Madhuri Barnwal, Mr. Baby, and Mr. Sibin for scheduling and technical assistance during scanning.

**Table 1**

| **Demographic information** | **Ratio / Mean ± SD** |
|---|---|
| Gender ratio (male: female) | ~3:1 (35:12) |
| Age (in years) | 21.70 ± 2.32 |
| PHQ-9 | 7.38 ± 4.30 |
| Fear Affect scale (anxiety) | 15.30 ± 4.82 |

Demographic information of the participants (*n* = 47). PHQ: Patient Health Questionnaire.

**Table 2. Networks and their associated functions, with PubMed search terms used for functional mapping within the two-system framework**

| Network | Cognitive, behavioural or physiological function mapped to the two-system framework | Search terms | Key references |
|---|---|---|---|
| Salience Network (SN) | interoception, autonomic activity, affective processing, attention | ("anxiety") AND ("salience network") AND ("interoception" OR "affective processing" OR "subjective anxiety feeling") | (Peters et al., 2016; Wu et al., 2019) |
| Default Mode Network (DMN) | anxious apprehension, emotion, memory-related processes | ("anxiety") AND ("functional magnetic resonance imaging" OR "fMRI") AND ("default mode network") AND ("anxious apprehension" OR "emotion" OR "affective processing" OR "memory-related processes") | (Baggio et al., 2025; Burdwood et al., 2016; Dixon et al., 2020, 2022; Pang et al., 2024; Rabany et al., 2017; Tao et al., 2015) |
| Frontoparietal Network (FPN) | attention, cognitive control, emotion-regulation | ("anxiety") AND ("functional magnetic resonance imaging" OR "fMRI") AND ("frontoparietal network") AND ("attention" OR "emotion regulation" | (Hakamata et al., 2018; Picó-Pérez et al., 2017a; van Oort et al., 2022) |
| | | OR "cognitive control" OR "memory-related processes" OR "subjective anxiety feeling") | |
| Visual Network (VN) | sensory-threat processing | ("anxiety") AND ("functional magnetic resonance imaging" OR "fMRI") AND ("Visual Cortex" OR "visual network" OR "occipital cortex") AND ("sensory-threat processing" OR "emotion" OR "threat-processing") | (Ding et al., 2011; Giménez et al., 2012; Liao et al., 2010; Sambuco et al., 2020; Sylvester et al., 2017) |
| Cerebellar Network (CN) | cognitive-affective evaluation, emotion regulation | ("anxiety") AND ("functional magnetic resonance imaging" OR "fMRI") AND ("Cerebellar Network" OR "cerebellum") AND ("sensory-threat processing" OR "emotion" OR "threat-processing" OR "cognition") | (Ko et al., 2025; Phillips et al., 2015; Picó-Pérez et al., 2017b) |

**Table 3. MRI Results.**

| **Analysis** | **ICA - Networks** | **Significant Voxels in Clusters** | **Cluster-size (*kE*)** | **Cluster-level *p*-FDR (*adjusted hypothesis-wide *p*FDR corrections across all five models / networks)** | **Effect size (T-value)** |
|---|---|---|---|---|---|
| Interaction Term: Behavioural response to Uncertainty of Threat × anxiety vs rsFC regression | Default Mode Network | MNI: +08, +38, +22 | 238 | 0.013 (0.039*) | $T_{(40)} = 5.39$ |
| | | Paracingulate Gyrus Right (MNI: +6, +38, +26) | 129 | | |
| | | Paracingulate Gyrus Left (MNI: -4, +38, +28) | 46 | | |
| | | Cingulate Gyrus, anterior division (MNI: +6, +36, +20) | 39 | | |
| | | Not labelled (MNI: +0, +38, +28) | 24 | | |
| | | | | | |
| | | MNI: +22, -80, -08<br><br>Occipital Fusiform Gyrus Right (MNI: +22, -82, -10)<br><br>Occipital Pole Right (MNI: +24, -96, -10) | 235<br><br>142<br><br>82 | 0.013 (0.039*) | $T_{(40)} = 5.04$ |
| Interaction Term: Physiological response to Uncertainty of Threat × anxiety vs rsFC regression | Salience Network | MNI: -10, -94, +30<br><br>Occipital Pole Left (MNI: -14, 94, +18) | 1252<br><br>545<br><br>187 | <0.001 (<0.001*) | $T_{(40)} = 5.58$ |
| | | Lateral Occipital Cortex, superior division Left (MNI: -22, -86, +22)<br><br>Intracalcarine Cortex Left (MNI: -8, -84, +6)<br><br>Cuneal Cortex left (MNI: -6, -86, +24)<br><br>Supracalcarine Cortex Left (MNI: -4, -88, +8)<br><br>Not labelled (MNI: -16, -86, +14) | 100<br><br>62<br><br>5<br><br>353 | | |
| Subjective response to anxiety vs rsFC regression | Cerebellar Network | MNI: -26, -74, +44<br><br>Lateral Occipital Cortex, superior division Left (MNI: -30, -68, +46)<br><br>Precuneous (MNI: -6, -64, +54) | 934<br><br>649<br><br>118 | 0.00001 (0.0001*) | $T_{(40)} = -6.00$ |
| | | Angular Gyrus Left (MNI: -50, -56, +38)<br><br>Superior Parietal Lobule Left (MNI: -22, -56, +54) | 70<br><br>38 | | |
| | | MNI: -38, +58, -06<br><br>Frontal Pole Left (MNI: -36, +52, -8)<br><br>Not labeled (MNI: -32, +50, -4) | 402<br><br>390<br><br>12 | 0.004 (0.016*) | $T_{(40)} = -5.25$ |
| | | MNI: +08, +10, +54<br><br>Supplementary Motor Cortex Right (MNI: +6, +2, +54)<br><br>Paracingulate Gyrus Right (MNI: +4, +12, +50)<br><br>Superior Frontal Gyrus Right (MNI: +6, +12, +54) | 295<br><br>142<br><br>47<br><br>15 | 0.015 (0.034*) | $T_{(40)} = 5.57$ |
| | | Supplementary Motor Cortex Left (MNI: -2, +6, +52) | 15 | | |
| | | Paracingulate Gyrus Left (MNI: -2, +6, +52) | 7 | | |
| | | Cingulate Gyrus, anterior division (MNI: +2, -4, +46) | 5 | | |
| | | Not labeled (MNI: +2, +8, +52) | 64 | | |

Note: ICA-to-cluster resting-state functional connectivity results surviving the initial cluster-level threshold ($p$FDR < 0.05) and the final $p$-values denoted with asterisks (*) indicate hypothesis-wide FDR corrections across all five models/networks for each independent hypothesis. FDR, false discovery rate; MNI, Montreal Neurological Institute coordinate system.

**Figure 1**

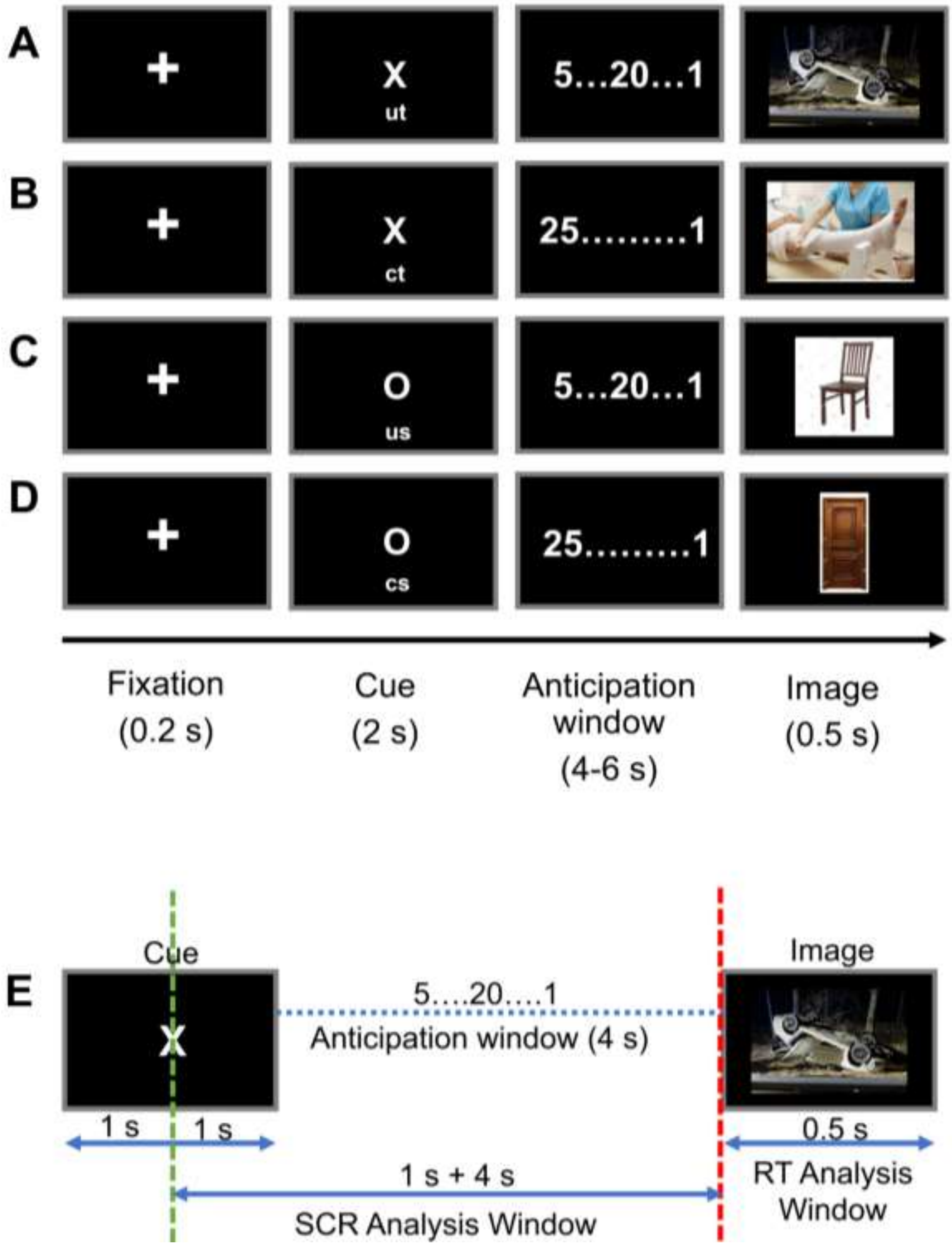

**Experiment Design.** (A–D) Each trial began with a fixation cross, followed by a cue indicating condition (X = Threat; O = Safety) presented for 2 s. Hints were presented below the cue to indicate the type of condition (ut = Uncertain Threat; us = Uncertain Safety; ct = Certain Threat; cs = Certain Safety). The cue was followed by an anticipation window lasting 4-6 s pseudorandomly, during which integers were displayed either randomly (Uncertain; A, C) or sequentially (Certain; B, D). At the end of the anticipation window, an image corresponding to the indicated condition (Threat: A, B; Safety: C, D) was presented for 0.5 s to elicit behavioural and physiological responses. The trial ended with a mouse click followed by the inter-trial interval. (E) Schematic illustrating the analysis windows for behavioural responses and phasic responses. Behavioural responses (reaction time; RT) were recorded from the end of the anticipation window until the mouse click. Phasic responses, recorded concurrently, were quantified from 1 s after cue onset (green line) to 4 s after the start of the anticipation window (red line). Images shown are illustrative and sourced from Shutterstock, as publication of IAPS images is prohibited.

**Figure 2**

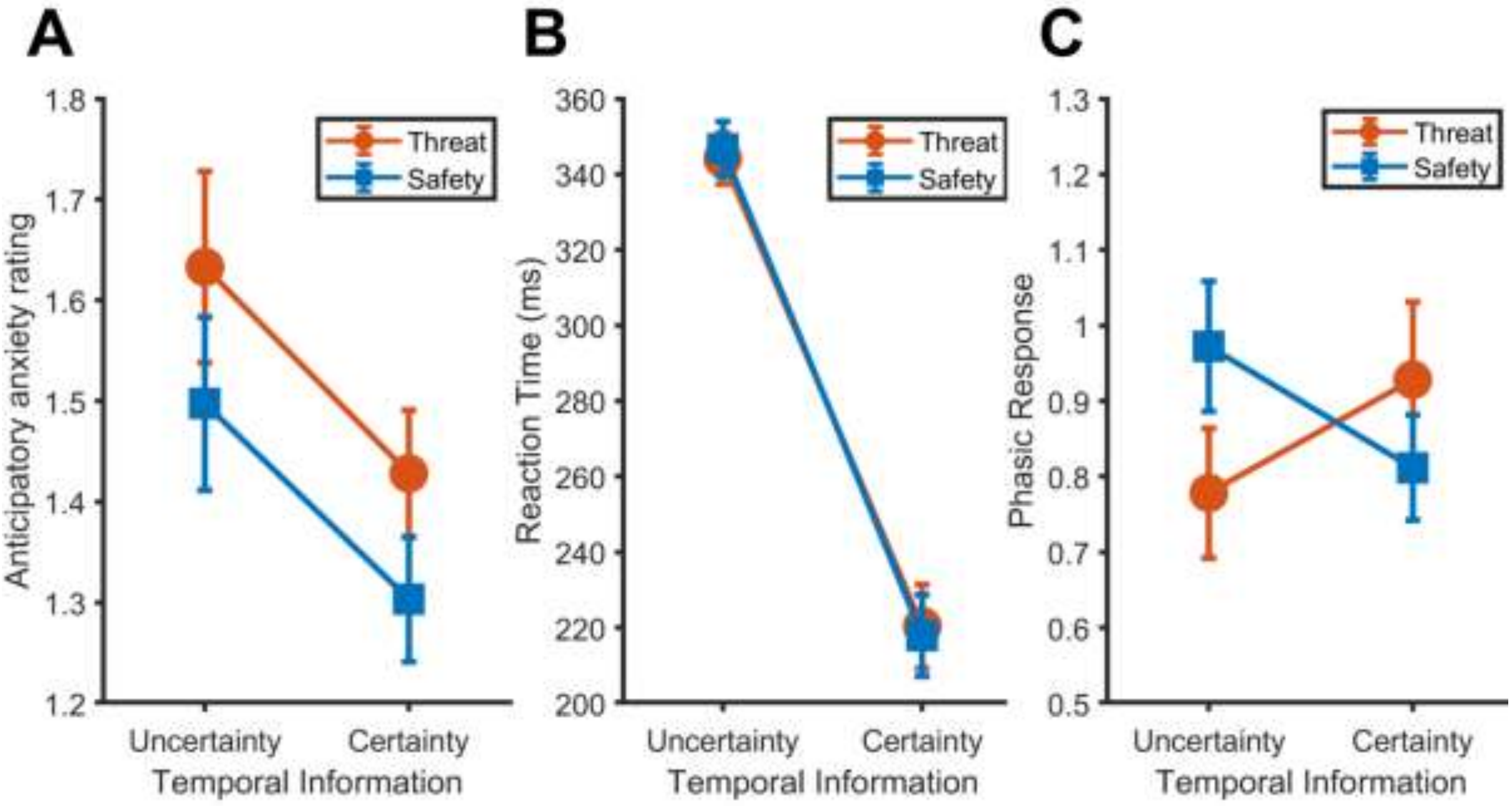

**Interaction plots illustrating task validation across – [A] anticipatory anxiety rating (subjective aversiveness), [B] reaction time, and [C] phasic response.** The circle ('threat' condition) and square ('safety') marker represent group mean and the error bars represent standard errors of the mean.

**Figure 3**

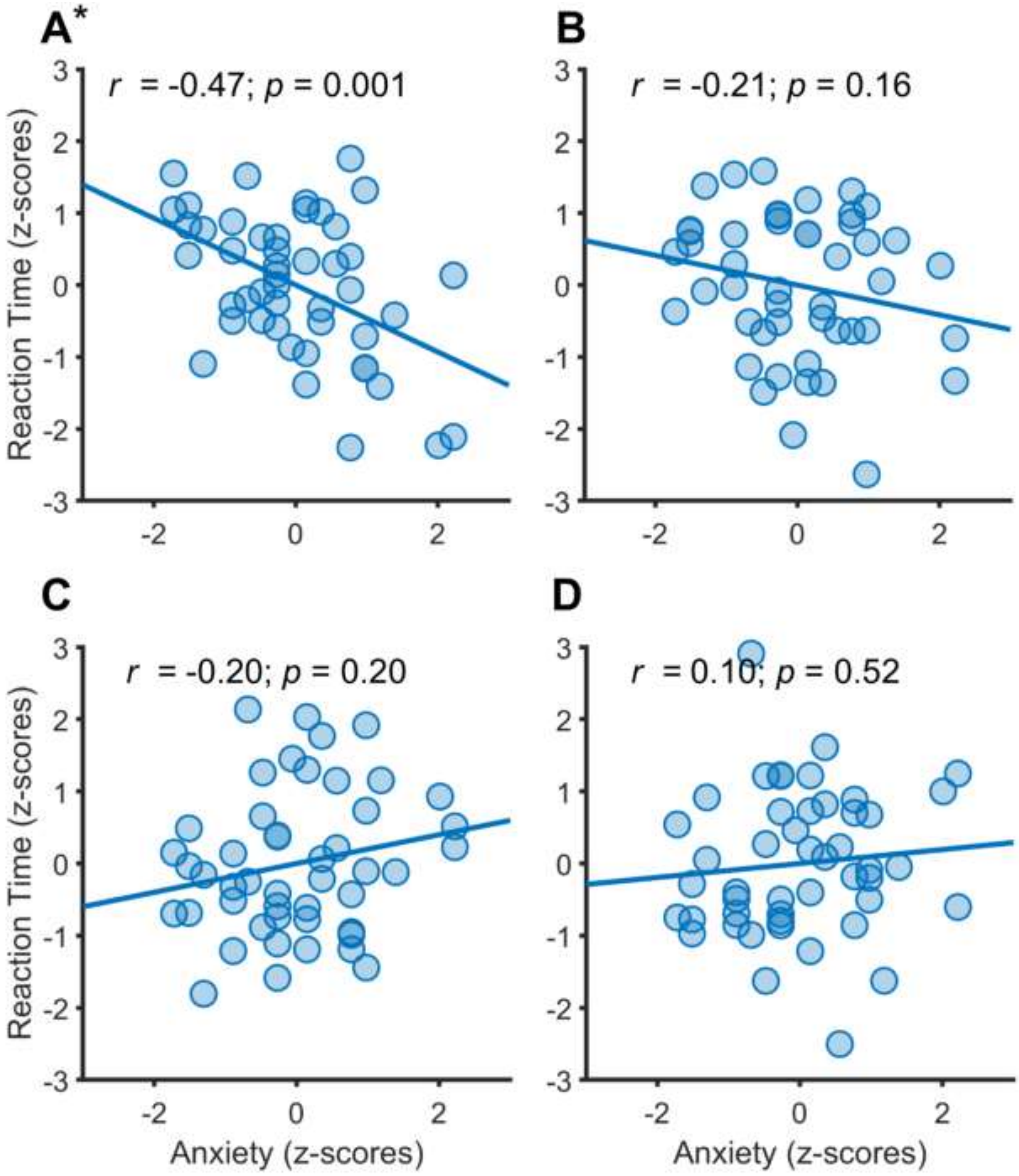

**Association between anxiety and behavioural response (A-D).** Each marker represents one participant, and the trend line represents the least square regression line for A) Uncertain Threat, B) Uncertain Safety, C) Certain Threat, and D) Certain Safety. **p* <0.05 for Uncertain Threat.

**Figure 4**

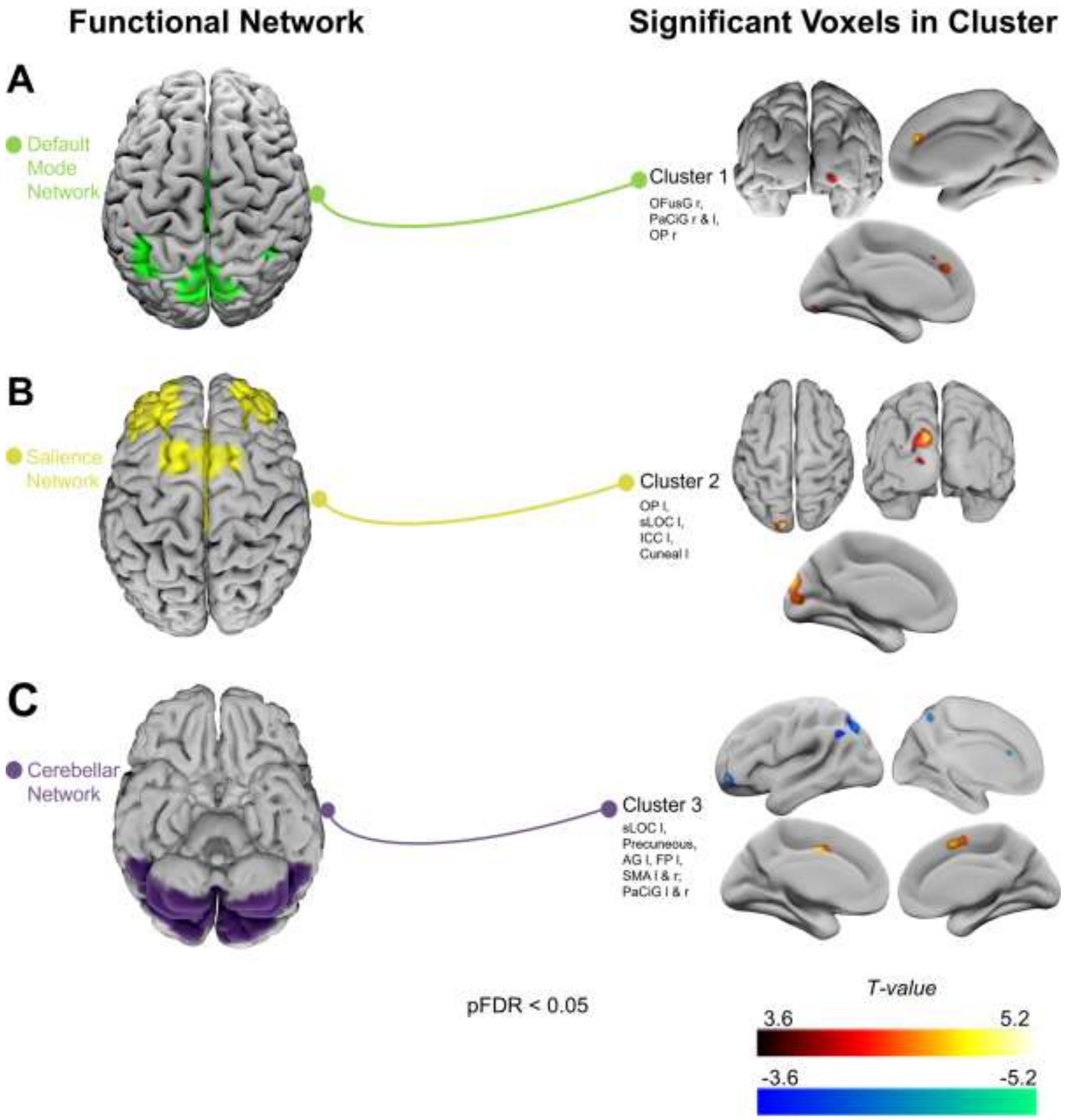

**Resting state functional connectivity between functional network and significant clusters linked to (A)** Anxiety modulated behavioural responses; **(B)** Anxiety modulated physiological responses; **(C)** Subjective conscious experience of anxiety after controlling for anxiety modulated behavioural and physiological responses. The warmer colours denote positive correlation between network and cluster whereas the cooler colours denote negative correlation. The visualisation of networks and significant clusters was done using Surf Ice (Rorden, 2025). OFusG r: Occipital Fusiform Gyrus right; PaCiG l & r: Paracingulate Gyrus left and right; OP r: Occipital Pole r: Occipital Pole right; OP l: Occipital Pole left; sLOC l:

Lateral occipital cortex, superior division left; ICC l: Intracalcarine Cortex left; Cuneal l: Cuneal left; AG l: Angular Gyrus Left; SMA l & r: Supplementary Motor Area left and right. FDR: False Discovery Rate.

**Title:** Intrinsic Brain Networks Underlying the Experience and Expression of Subclinical Anxiety

**Authors:** Shruti Kinger, Naviya Lall, Mrinmoy Chakrabarty

## Supplementary Information

**S1: *Representative sample of skin conductance response (SCR) analysis***

Figure S1 illustrates the AUC calculation for one session containing a total of 32 trials from a single participant. The AUC was computed for each correct trial from the time point at which the phasic component first exceeded 0.01 until it fell below threshold of 0.01 across the anticipation window of 5 s. Trials with an AUC of zero correspond to those in which the SCR did not exceed threshold at any point during the anticipation window.

**Figure S1**

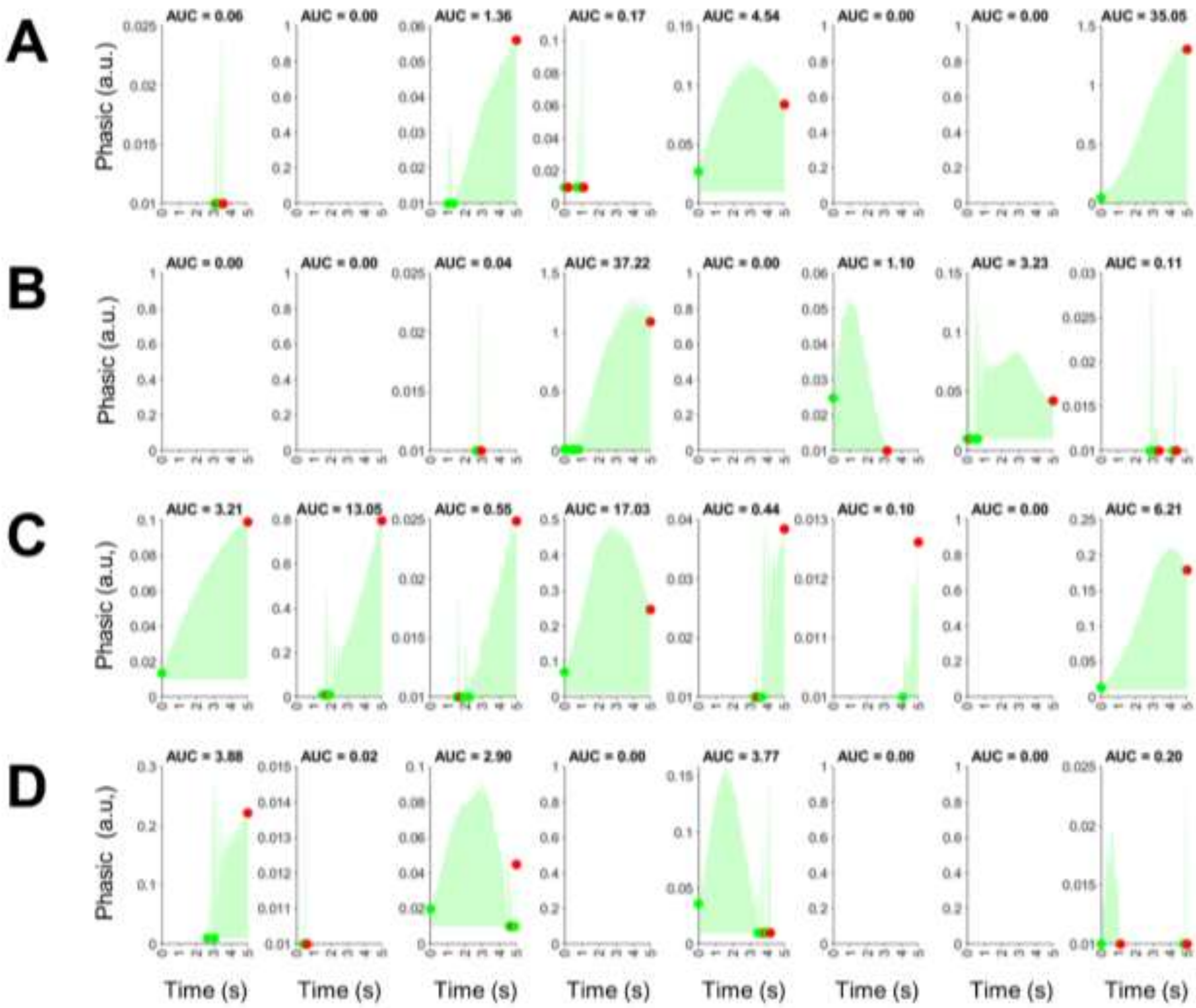

**Sample plot depicting the calculation of the area under the curve (AUC).** Each row represents the four experimental conditions – (A) Uncertain Threat, (B) Uncertain Safety, (C) Certain Threat, (D) Certain Safety. Each column represents one trial. The green solid circle represents the starting point and the red solid circle represents the end point.

**S2: *Comparison of Mean Absolute Deviation: Repository Data vs. Present Study***

To ensure that the data in the present study were not noisier than expected, we compared them with a repository dataset (Lor et al., 2025). First, the AUC for the repository dataset was computed using the same procedure described in the Methods section (Skin Conductance Data Analysis). Second, participants with an AUC of zero in the repository dataset were excluded, leaving 29 participants for analysis. Third, trials with an AUC of zero were excluded from the analysis for each participant. Fourth, trials across the four experimental conditions of the present study were merged for each participant, and the least common denominator in the number of trials of each participant was extracted to enable comparison with the repository data. Fifth, the mean absolute deviation across trials was computed for each participant for both the repository and the primary dataset. Finally, after confirming normality using Lilliefors tests (all *p*s ≈ 0.5), mean absolute deviation was compared between the repository and primary datasets using an independent t-test. This comparison indicated that the data from the present study were not noisier than those from the repository dataset ($t = -1.62$, degrees of freedom = 56, 95% confidence interval = [-1, 0.1], $p = 0.11$)

**S3: Skin Conductance Results**

Figure S2 depicts the correlation between anxiety scores and the phasic component pertaining to four conditions – Uncertain Threat, Uncertain Safety, Certain Threat, Certain Safety.

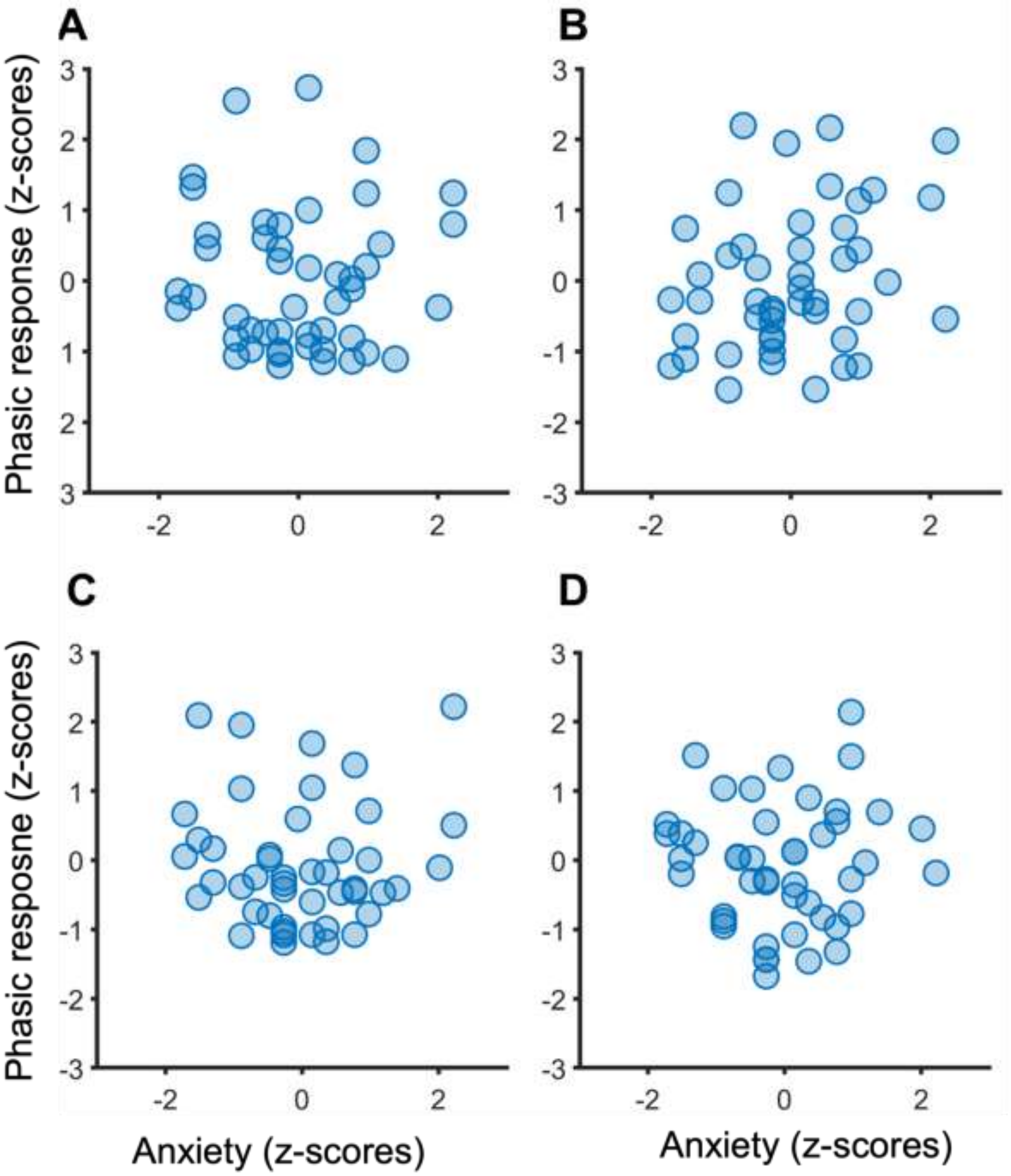

**Figure S2: Association between anxiety and physiological response.** Each marker represents one participant for A) Uncertain Threat, B) Uncertain Safety, C) Certain Threat, and D) Certain Safety.

**S4**: **Mean time-course of phasic component across participants.**

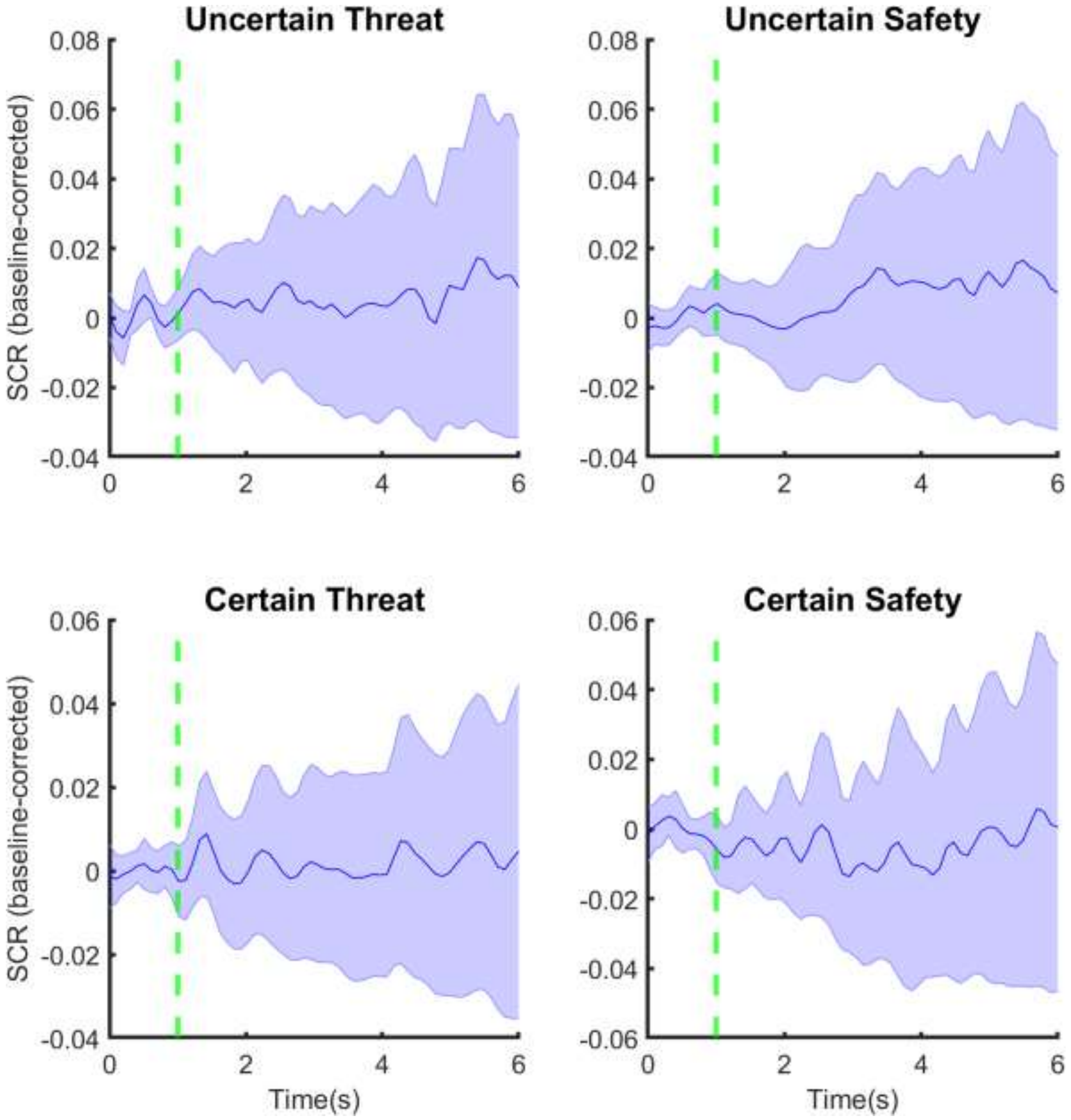

**Figure S3:** Mean time-course of baseline-corrected phasic component (Skin Conductance Response or SCR) across participants from baseline (0 ms) to the end of anticipatory period: A) Uncertain Threat, B) Uncertain Safety, C) Certain Threat, and D) Certain Safety. The green dotted vertical line indicates stimulus onset. Shaded error patches represent the 95% confidence interval around the mean, calculated from the standard error of the mean (SEM) and plotted using the 'shadedErrorBar' MATLAB function.

**S5: Networks overlaid on average T1-weighted template.**

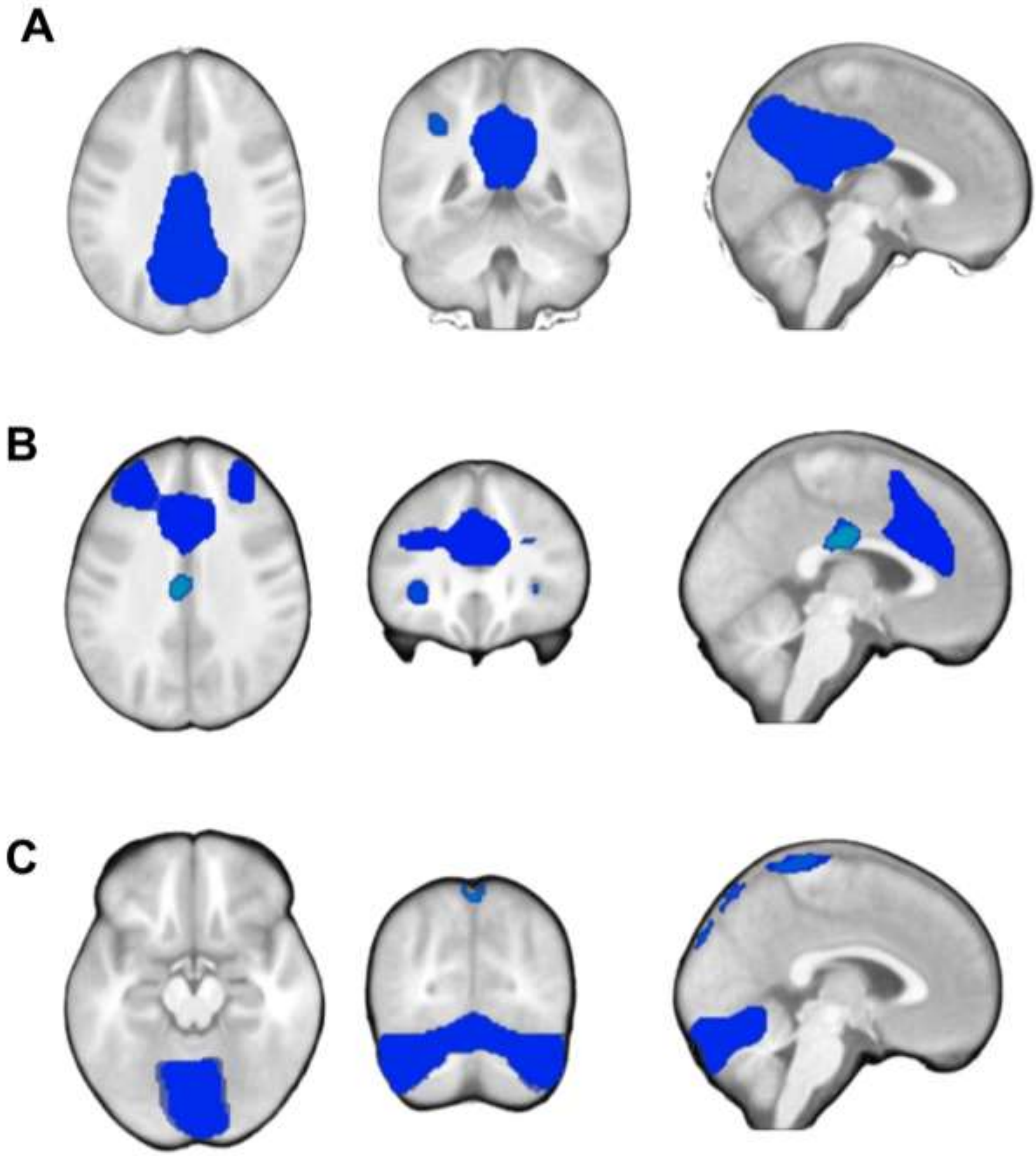

**Figure S4:** Default Mode Network (A), Salience Network (B), and Cerebellar Network (C) overlaid on averaged normalised T1-weighted template obtained from data (scans) acquired from 47 participants.

**S6: Networks overlaid on average T2-weighted template.**

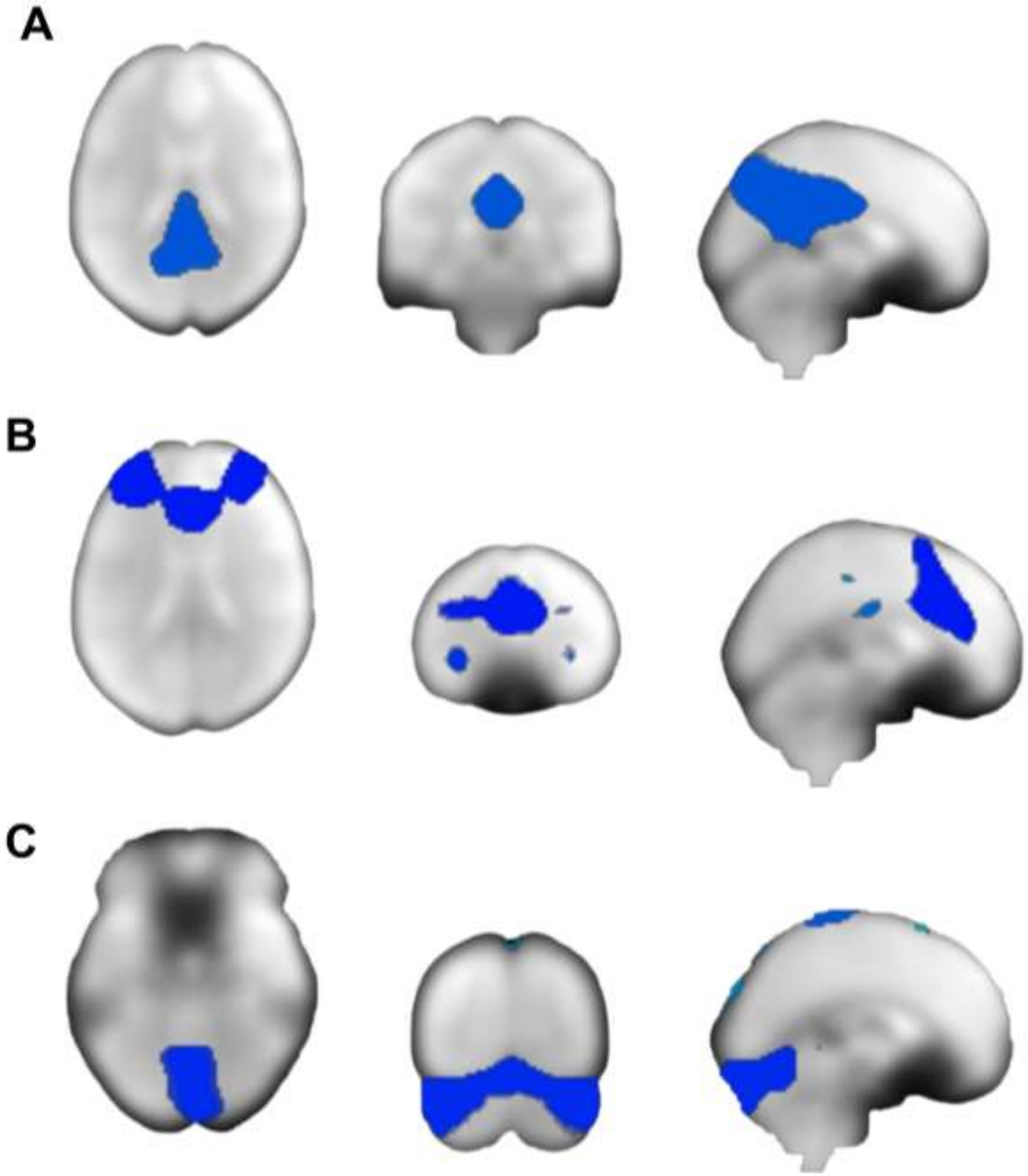

**Figure S5:** Default Mode Network (A), Salience Network (B), and Cerebellar Network (C) overlaid on average T2-weighted (denoised and smoothed) template obtained from data (scans) acquired from 47 participants.

**S7: Association between anxiety-modulated behavioural response specific to uncertain threat and resting state functional connectivity**

| **Analysis** | **ICA - Networks** | **Significant Voxels in Clusters** | **Cluster-size (*kE*)** | **Cluster-level *p*-FDR (*adjusted hypothesis-wide *p*FDR corrections across all five models / networks)** | **Effect size (T-value)** |
|---|---|---|---|---|---|
| Interaction Term: Behavioural response to Uncertainty of Threat × anxiety vs rsFC regression | Default Mode Network | MNI: -02, +30, +36<br><br>Paracingulate Gyrus Right (MNI: +6, +30, +34)<br><br>Paracingulate Gyrus Left (MNI: -4, +28, +38)<br><br>Superior Frontal Gyrus Left (MNI: -6, +32, +42)<br><br>Not labelled (MNI: +0, +30, +38) | 337<br><br>181<br><br>99<br><br>5<br><br>52 | 0.003 (0.015*) | $T_{(40)} = 5.03$ |

**S8: Anxiety moderated association between aversiveness to uncertainty and resting state functional connectivity or rsFC**

We examined whether anxiety moderated the association between aversiveness rating and rsFC to determine whether the relationship between aversiveness rating and rsFC varied as a function of anxiety levels. The moderation analysis revealed a significant positive association between the anxiety-modulated aversiveness rating and rsFC (Figure S6; Table S1), indicating that the relationship between aversiveness rating and rsFC became increasingly positive with higher levels of anxiety. Significant rsFC was observed between the default mode network and a cluster within the visual cortex encompassing the left occipital pole, left inferior and superior division of the lateral occipital cortex.

Given the role of the default mode network in memory and self-related processing and the involvement of the visual cortex in visual processing (Azarias et al., 2025; Chadick & Gazzaley, 2011; Wysiadecki et al., 2021), this increased connectivity may suggest that prior aversive experiences influence the engagement of visual regions in monitoring for potential aversive events.

**Table S1.** MRI Results

| **Analysis** | **ICA - Networks** | **Significant Voxels in Clusters** | **Cluster-size (*kE*)** | **Cluster-level *p*-FDR/ adjusted hypothesis-wide correction across all regression models** | **Effect size (T-value)** |
|---|---|---|---|---|---|
| Interaction Term: Anticipatory Anxiety Rating to Uncertainty (Subjective aversiveness) of Threat × anxiety vs rsFC regression | Default Mode Network | MNI: -30, -96, +10<br><br>Occipital Pole Left<br>(MNI: -30, -94, +8)<br><br>Lateral occipital cortex, inferior division Left<br>(MNI: -42, -86, +2)<br><br>Lateral occipital cortex, superior division Left<br>(MNI: -42, -88, +14)<br><br>Not-labelled | 350<br><br>195<br><br>81<br><br>11<br><br>63 | 0.003<br>(0.015*) | $T_{(40)} = 5.43$ |
| | | (MNI: +0, +38, +28) | | | |

Note: ICA-to-cluster resting-state functional connectivity results surviving the initial cluster-level threshold (*p*FDR < 0.05) and the final *p*-values denoted with asterisks (*) indicate hypothesis-wide FDR corrections across all five models/networks for each independent hypothesis. FDR, false discovery rate; MNI, Montreal Neurological Institute coordinate system.

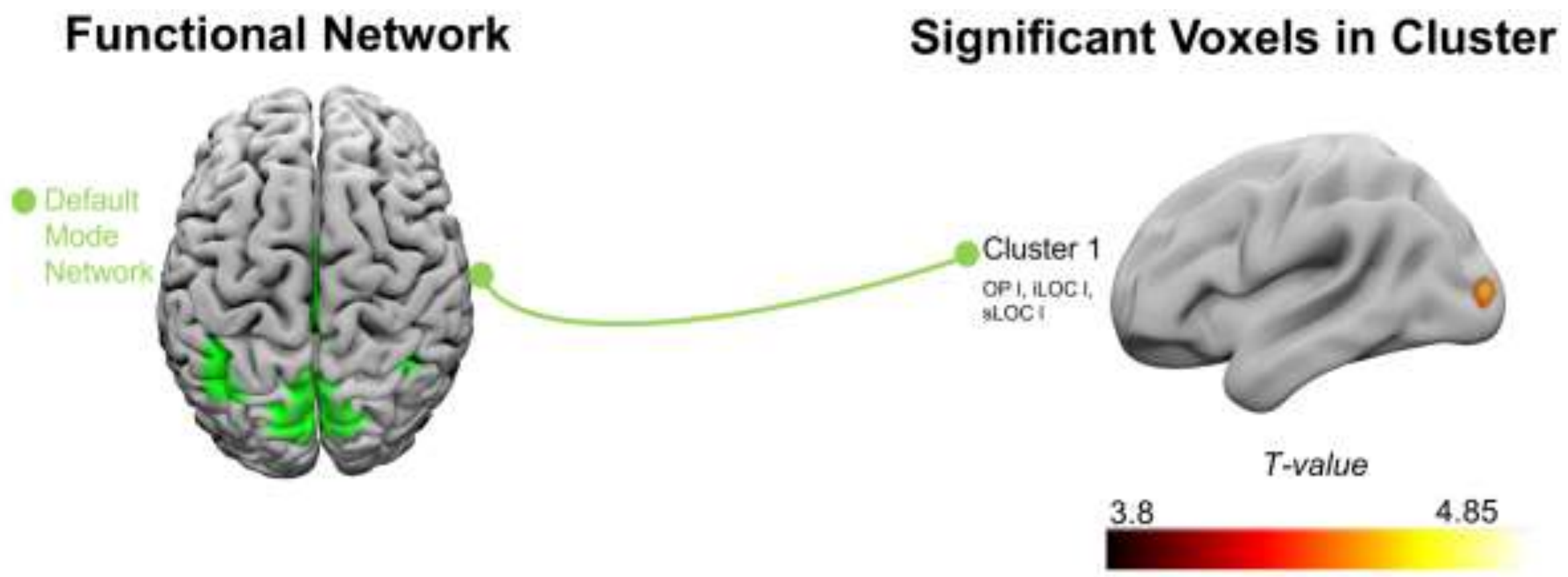

**Figure S6. Resting state functional connectivity between functional network and significant clusters linked to anxiety modulated aversiveness rating**. The warmer colours denote positive correlation between network and cluster. The visualisation of networks and significant clusters was done using Surf Ice (Rorden, 2025). OP l: Occipital Pole left; iLOC l: Lateral occipital cortex, inferior division left; sLOC l: Lateral occipital cortex, superior division left